\documentclass[usenatbib]{mn2e}
\usepackage[utf8]{inputenc} 
\pdfoutput=1 
\usepackage{graphicx}
\usepackage{subfigure}
\usepackage{amsmath}
\usepackage[english]{babel}
\usepackage{natbib}
\usepackage{longtable}
\usepackage{lscape}

\bibpunct{(}{)}{;}{a}{}{,}
\newcommand{\vect}[1]{\ensuremath{\mathbf{#1}}}

\newcommand{\bv}{\mathbf{v}}
\newcommand{\bb}{\mathbf{B}}
\newcommand{\mrho}{\overline{\varrho}}
\newcommand{\mT}{\overline{T}}\newcommand{\mP}{\overline{P}}
\newcommand*{\dif}{\mathop{}\!\mathrm{d}}
\newcommand{\Prandtl}{\ensuremath{Pr}}
\newcommand{\Pm}{\ensuremath{Pm}}
\newcommand{\Pmc}{\ensuremath{Pm_\text{c}}}
\newcommand{\Rm}{\ensuremath{Rm}}
\newcommand{\Rmc}{\ensuremath{Rm_\text{c}}}
\newcommand{\Ro}{\ensuremath{Ro}}
\newcommand{\Ekman}{\ensuremath{E}}
\newcommand{\Rol}{\ensuremath{Ro_\ell}}
\newcommand{\Rolr}{\ensuremath{Ro_\ell^\star}}
\newcommand{\lc}{\ensuremath{\ell_c}}
\newcommand{\Roz}{\ensuremath{Ro_z}}
\newcommand{\Ra}{\ensuremath{Ra}}
\newcommand{\Rac}{\ensuremath{Ra_\text{c}}}

\newcommand{\fdipAX}{\ensuremath{{f_\text{dip}}_\text{ax}}}
\newcommand{\Lo}{\ensuremath{Lo}}

\newcommand{\Nrho}{\ensuremath{N_\varrho}}

\newcommand{\Ek}{\ensuremath{E_\mathrm{k}}}
\newcommand{\Em}{\ensuremath{E_\mathrm{m}}}
\newcommand{\ro}{\ensuremath{r_{\!o}}}
\newcommand{\ri}{\ensuremath{r_{i}}}
\newcommand{\aspectratio}{\ensuremath{\chi}}
\newcommand{\mm}{\ensuremath{\mathrm{\texttt{m}}}}
\newcommand{\dd}{\ensuremath{\mathrm{\texttt{d}}}}

\newlength{\mafig}
\newlength{\mafigME}
\newlength{\figsl}
\title[]{Dipolar dynamos in stratified systems}
\author[]{R. Raynaud$^{1,2}$\thanks{E-mails: raphael.raynaud@ens.fr
    (RR); ludovic@lra.ens.fr (LP); dormy@phys.ens.fr (ED) },
  L. Petitdemange$^{1,2}$\footnotemark[1] and
  E. Dormy$^{1,3}$\footnotemark[1]\\ $^{1}$MAG (ENS/IPGP), LRA,
  D\'epartement de Physique, \'Ecole normale sup\'erieure, 24 rue
  Lhomond, 75252 Paris Cedex 5, France\\ $^{2}$LERMA, Observatoire de
  Paris, PSL Research University, CNRS, UMR 8112, F-75014, Paris,
  France \\ $^{3}$IPGP, CNRS UMR 7154, 75005 Paris, France}
\begin{document}

\date{}

\pagerange{\pageref{firstpage}--\pageref{lastpage}} \pubyear{2014}

\def\LaTeX{L\kern-.36em\raise.3ex\hbox{a}\kern-.15em
    T\kern-.1667em\lower.7ex\hbox{E}\kern-.125emX}

\maketitle

\label{firstpage}

\begin{abstract}
  Observations of low-mass stars reveal a variety of magnetic field
  topologies ranging from large-scale, axial dipoles to more complex
  magnetic fields. At the same time, three-dimensional spherical
  simulations of convectively driven dynamos reproduce a similar
  diversity, which is commonly obtained either with Boussinesq models
  or with more realistic models based on the anelastic approximation,
  which take into account the variation of the density with depth
  throughout the convection zone. Nevertheless, a conclusion from
  different anelastic studies is that dipolar solutions seem more
  difficult to obtain as soon as substantial stratifications are
  considered. In this paper, we aim at clarifying this point by
  investigating in more detail the influence of the density
  stratification on dipolar dynamos. To that end, we rely on a
  systematic parameter study that allows us to clearly follow the
  evolution of the stability domain of the dipolar branch as the
  density stratification is increased.  The impact of the density
  stratification both on the dynamo onset and the dipole collapse is
  discussed and compared to previous Boussinesq results. Furthermore,
  our study indicates that the loss of the dipolar branch does not
  ensue from a specific modification of the dynamo mechanisms related
  to the background stratification, but could instead result from a
  bias as our observations naturally favour a certain domain in the
  parameter space characterized by moderate values of the Ekman
  number, owing to current computational limitations.  Moreover, we
  also show that the critical magnetic Reynolds number of the dipolar
  branch is scarcely modified by the increase of the density
  stratification, which provides an important insight into the global
  understanding of the impact of the density stratification on the
  stability domain of the dipolar dynamo branch.
\end{abstract}


\section{Introduction}

Observations of low-mass stars reveal very different magnetic field
topologies, ranging from small-scale fields to large-scale dipolar
fields, and the last advances in spectropolarimetry should enable one
to improve the understanding of the magnetic fields of solar-type
stars \citep{donati09,morinD2010}. Among the three suggestions
advanced by Larmor to explain the generation of such magnetic fields
\citep{larmor1919}, it is now the consensus that their decay is
prevented by the action of self-excited dynamos induced by the
turbulent motions that occur in stellar interiors. More often, these
motions are assumed to be driven by convection, owing to the
temperature difference between the inner core and the cooler
surface. In dynamo theory, this partial transfer of the kinetic energy
of a conducting fluid into magnetic energy is an instability process:
above a certain threshold, electrical currents start to be amplified
by the fluid flow, so that a magnetic field can be sustained against
the resistive decay due to ohmic dissipation.

After \cite{glatzroberts95}, numerical modelling of self-consistent
dynamos underwent considerable development (in contrast with the small
number of successful experimental studies). However, despite the
continuous increase of computer power, direct numerical simulations
still face the difficulty to resolve a vast range of spatial and
temporal scales when attempting to simulate a three-dimensional
turbulent flow on a magnetic diffusion time-scale. As a
simplification, one usually resorts to some convective approximations,
and most of the early studies were relying on the Boussinesq
approximation, which performs well as long as variations in pressure
hardly affect the density of the fluid. However, this assumption is
not valid to describe convection in large stratified systems such as
stars or gas giants, in which the density typically varies over many
scale-heights between the top and bottom of the convection zone. This
limitation of the Boussinesq approximation is basically what motivated
the use of the anelastic approximation, originally developed to study
atmospheric convection \citep{ogura62,gough69}, to model convection in
the Earth core and stellar interiors. Indeed, if we assume that the
overall system remains close to an adiabatically stratified reference
state at marginal stability so that convective motions can be treated
as small perturbations (which in turns implies that typical velocities
remain small compare to the speed of sound), then the anelastic
approximation allows us to take some stratification into account while
filtering out sound waves for faster numerical integration.  This
approximation can be found in the literature under slightly different
formulations \citep{gilman81,braginsky95,lantz99,anufriev2005,
  berkoff2010,jones11,alboussiere13}, which are in part compared in
\cite{brown12}.

Just as in Boussinesq models
\citep{christensen06,schrinner12,yadav13}, magnetic fields obtained
in anelastic simulations \citep{gastine12,duarte2013,schrinner2014}
fall into two categories: dipolar dynamos, dominated by a large-scale
axial dipole component, and multipolar dynamos, characterized by a
more complex field topology with higher spatial and temporal
variability. However, these studies identified several differences
specific to anelastic dynamos. For instance, dipolar solutions seem
more difficult to obtain as the density stratification is increased
\citep{gastine12,jones2014}.  We found in \cite{schrinner2014} that
for a given \Nrho{}, \Ekman{} and \Prandtl{}, there seems to exist a
critical magnetic \Pmc{} below which the dipolar solution is not
stable, and the higher the density stratification, the higher this
critical magnetic Prandtl number. Furthermore, multipolar dynamos with
a magnetic field configuration dominated by an equatorial dipole seem
more easily realized with anelastic models than with Boussinesq
models. However, we show in \citet{raynaud2014} that this
characteristic also stands for weakly stratified models, since it is
actually related to the use of different mass distributions. Indeed,
the gravity profile may strongly influence the localization of the
convective cells, depending on whether one considers a homogeneous ($g
\propto r$) or a central mass ($g \propto 1/r^2$) distribution: as
opposed to the former, the latter results in the concentration of the
convective cells close to the inner sphere, which favours the
emergence of a less diffusive large-scale $m=1$ mode at the outer
surface of the model.

Our last study of weakly stratified models with a central mass
distribution naturally constitutes an appropriate reference basis from
which a detailed understanding of the role of the density
stratification in anelastic dynamo models can be achieved. In this
paper, we will primarily focus on dipolar dynamos. We aim at
clarifying apparent contradictions between previous anelastic studies
by investigating in more detail the evolution of the stability domain
of the dipolar branch when increasing the density stratification. To
that end, we rely on a systematic parameter study consisting of 119
three-dimensional, self-consistent dynamo models obtained by direct
numerical simulations. As opposed to previous studies that were
focusing on Jupiter's magnetic field
\citep{duarte2013,gastine2014,jones2014}, we do not consider here more
realistic models to reproduce a particular observation, but instead
try to understand systematic and general tendencies in anelastic
models, as a function of the physical control parameters. The
anelastic equations are recalled in Section~\ref{s:eq} and we present
our results in Section~\ref{s:results}. The complete list of numerical
simulations performed for this study is given in Table~\ref{table}
(see Appendix~\ref{s:appendix}).
\begin{figure*}
  \centering
  \subfigure[$\Nrho=0.5$]{
    \includegraphics[width=\mafig]{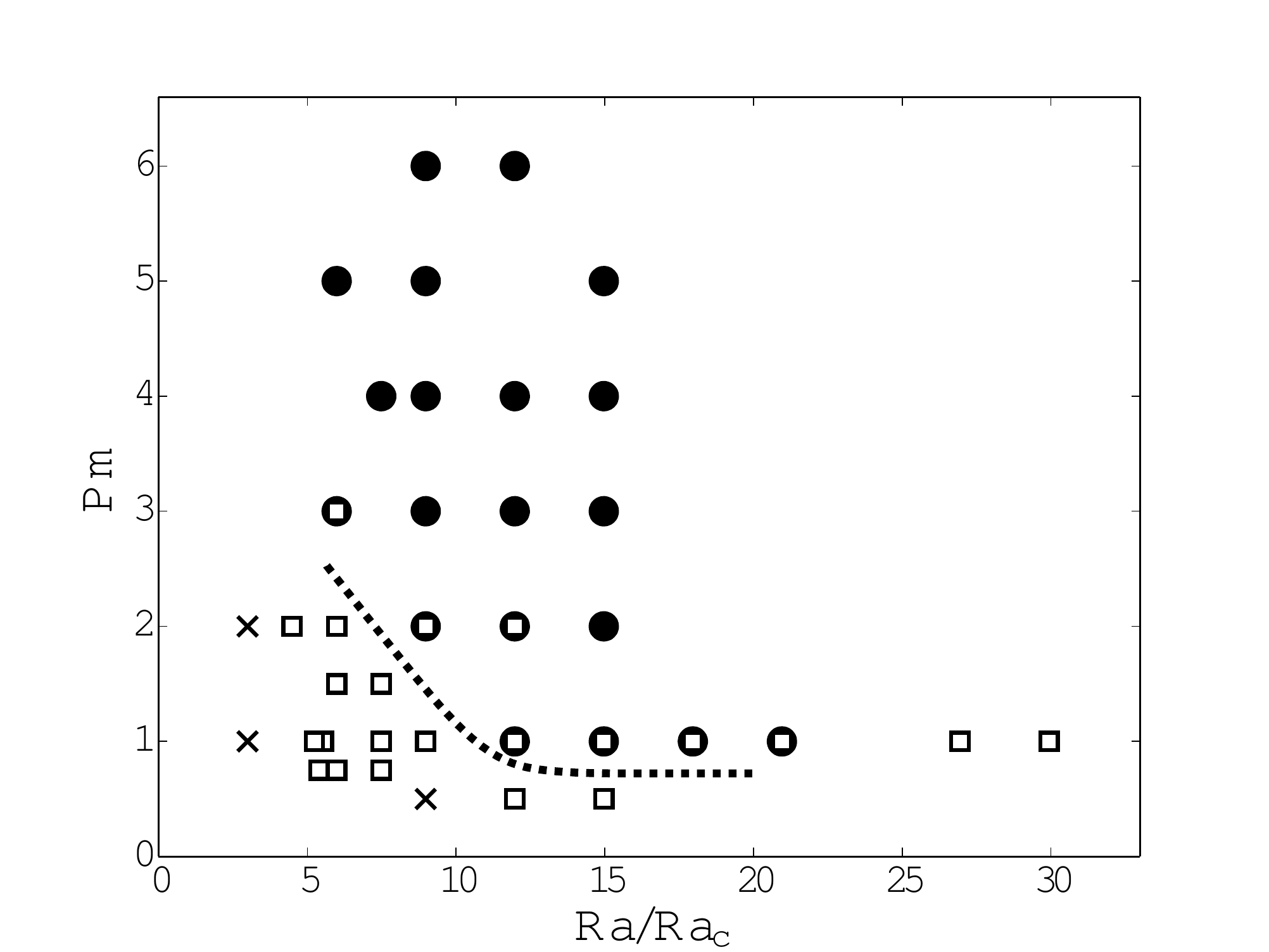}\label{sf:05}}
  \subfigure[$\Nrho=0.5$]{
    \includegraphics[width=\mafig]{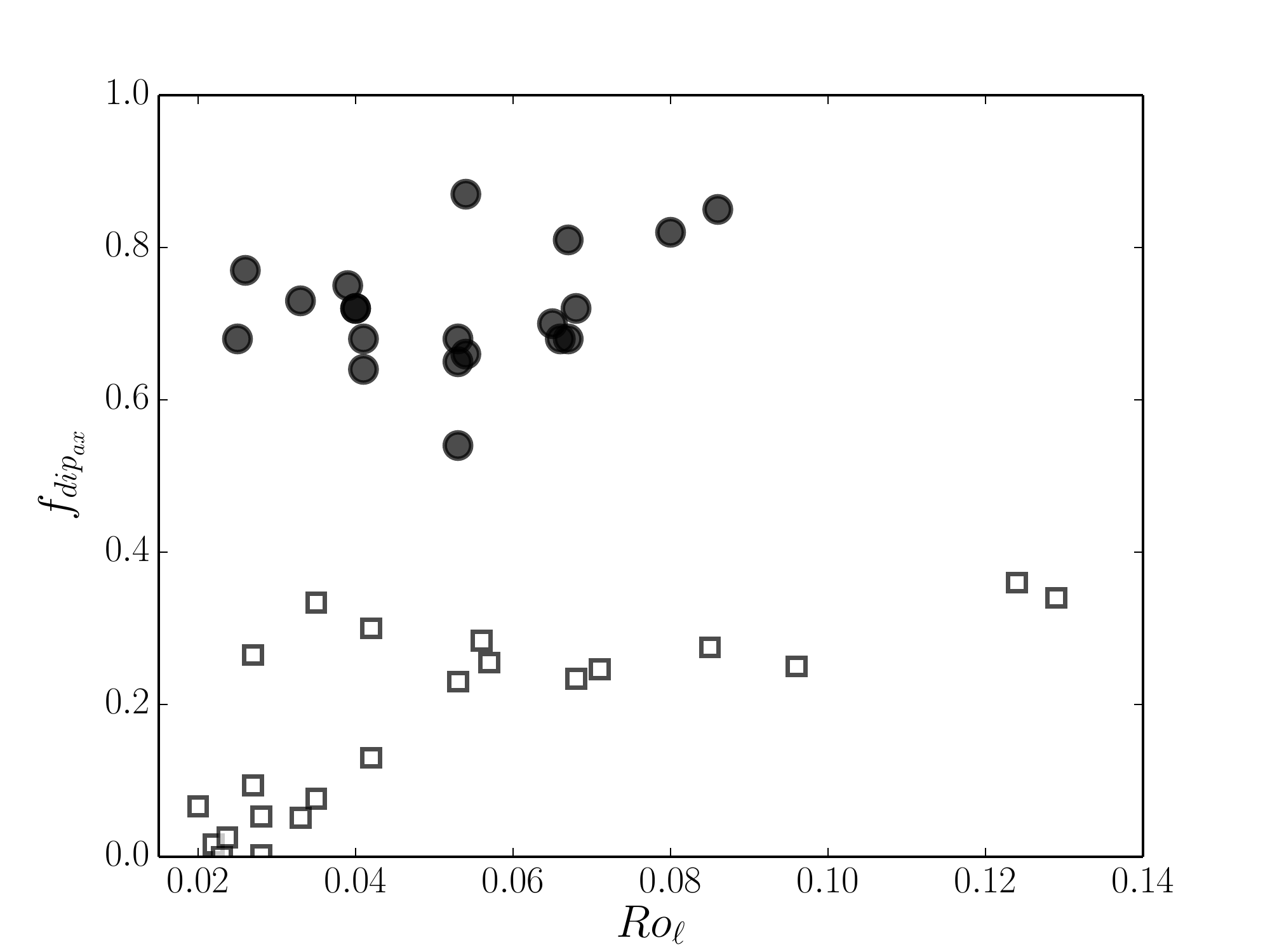}\label{sf:fdip05}}

  \subfigure[$\Nrho=1.5$]{
    \includegraphics[width=\mafig]{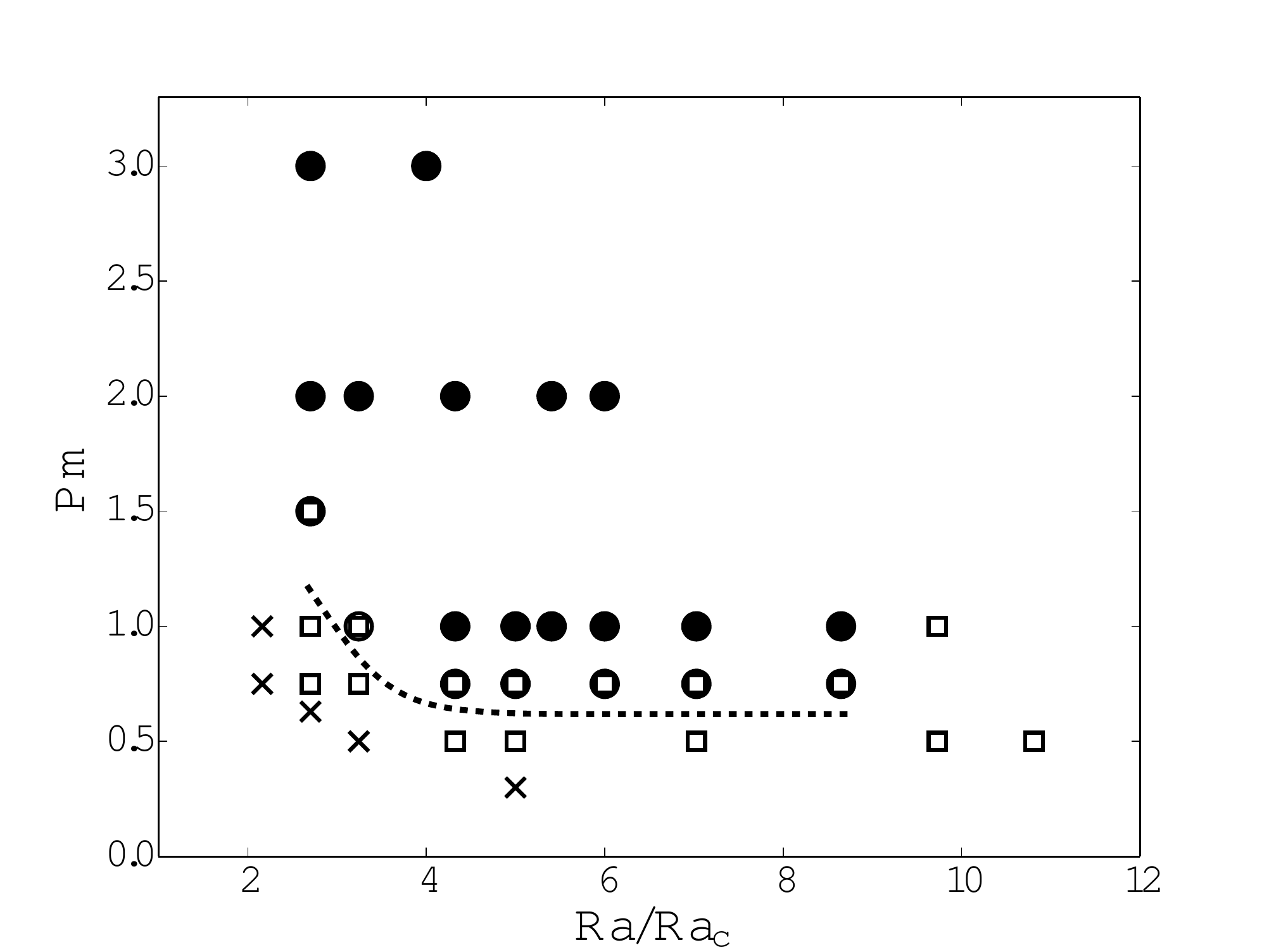}\label{sf:15}}
  \subfigure[$\Nrho=1.5$]{
    \includegraphics[width=\mafig]{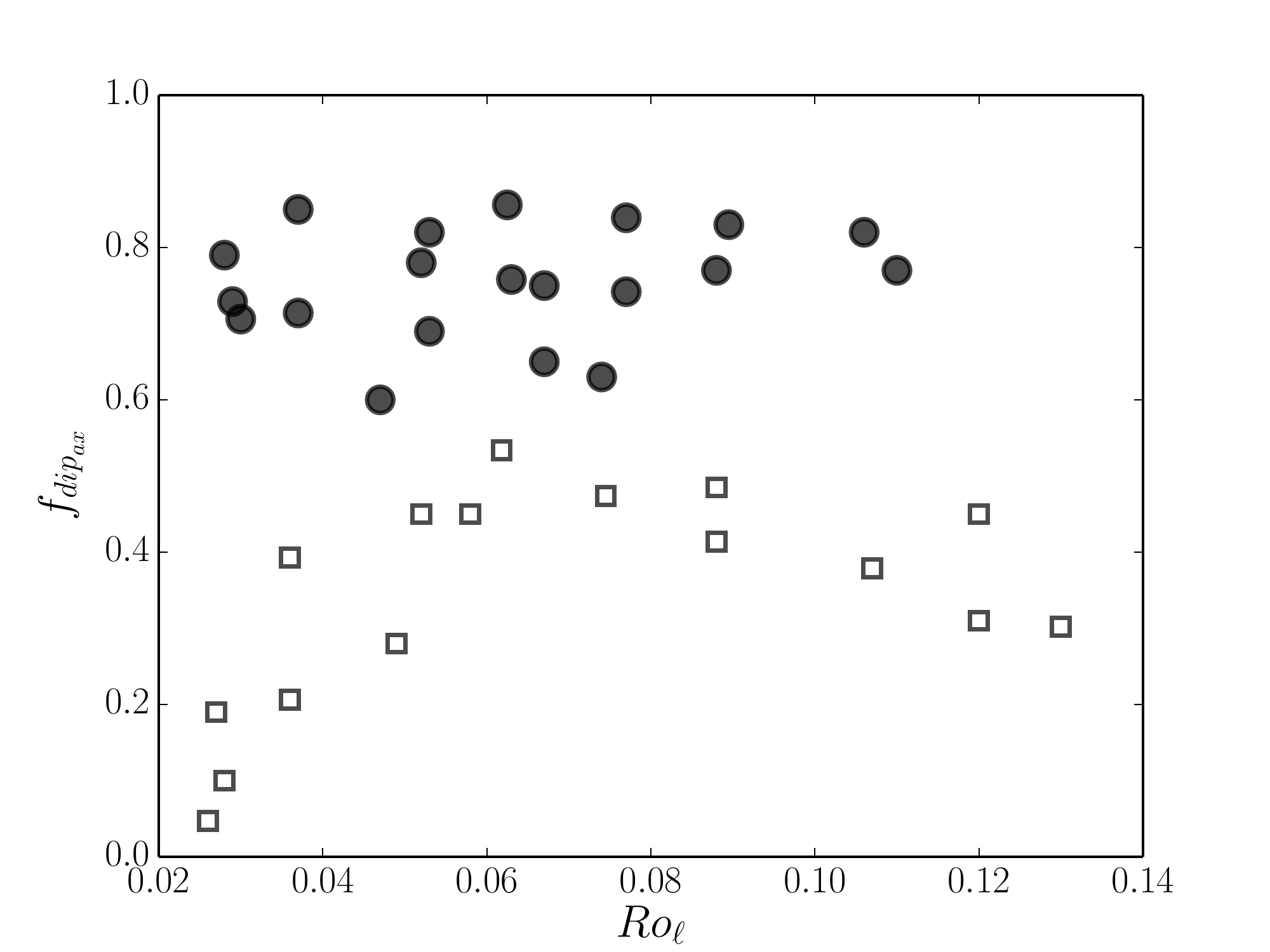}\label{sf:fdip15}}

  \subfigure[$\Nrho=2.0$]{
    \includegraphics[width=\mafig]{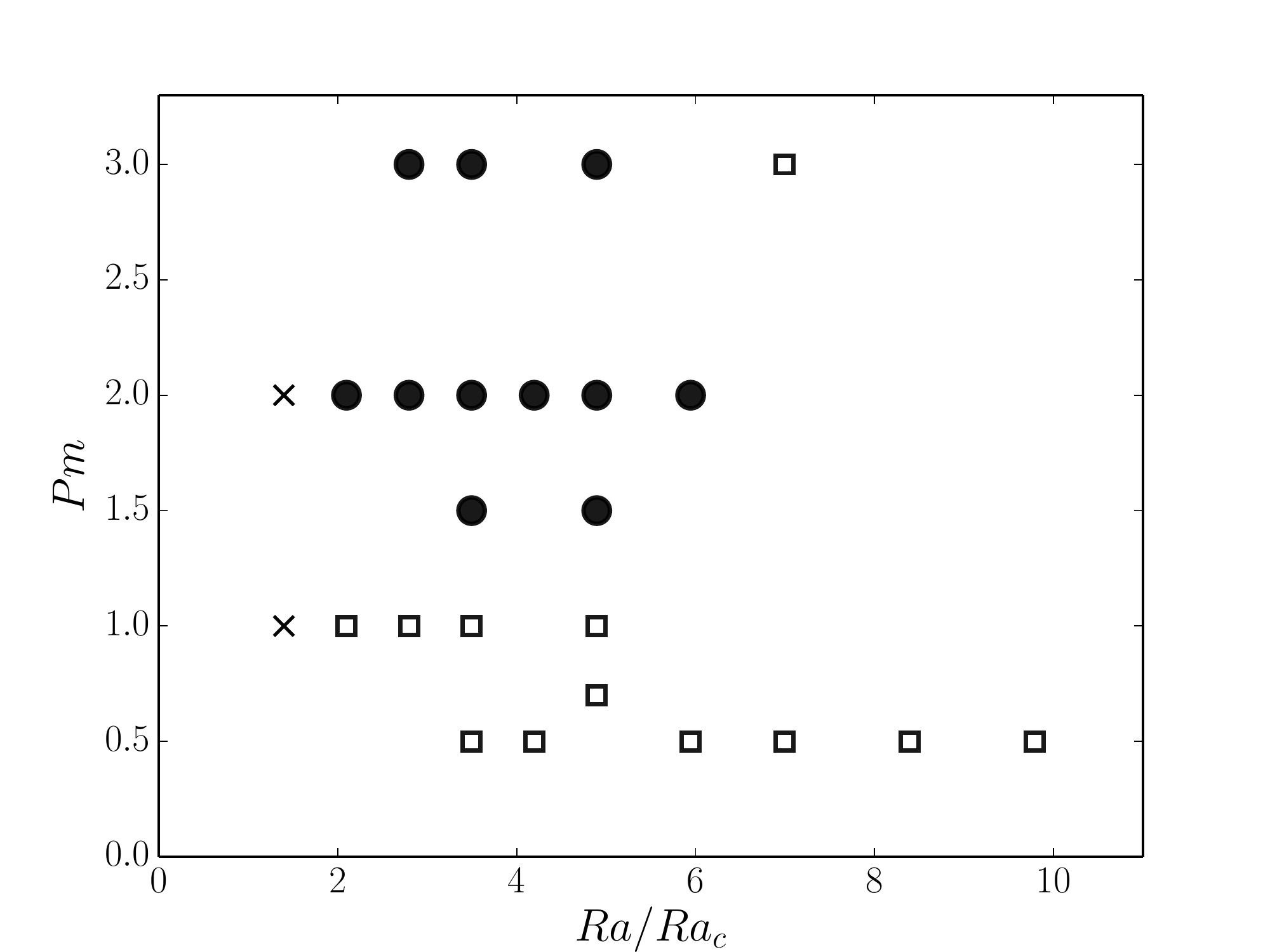}\label{sf:20}}
  \subfigure[$\Nrho=2.0$]{
    \includegraphics[width=\mafig]{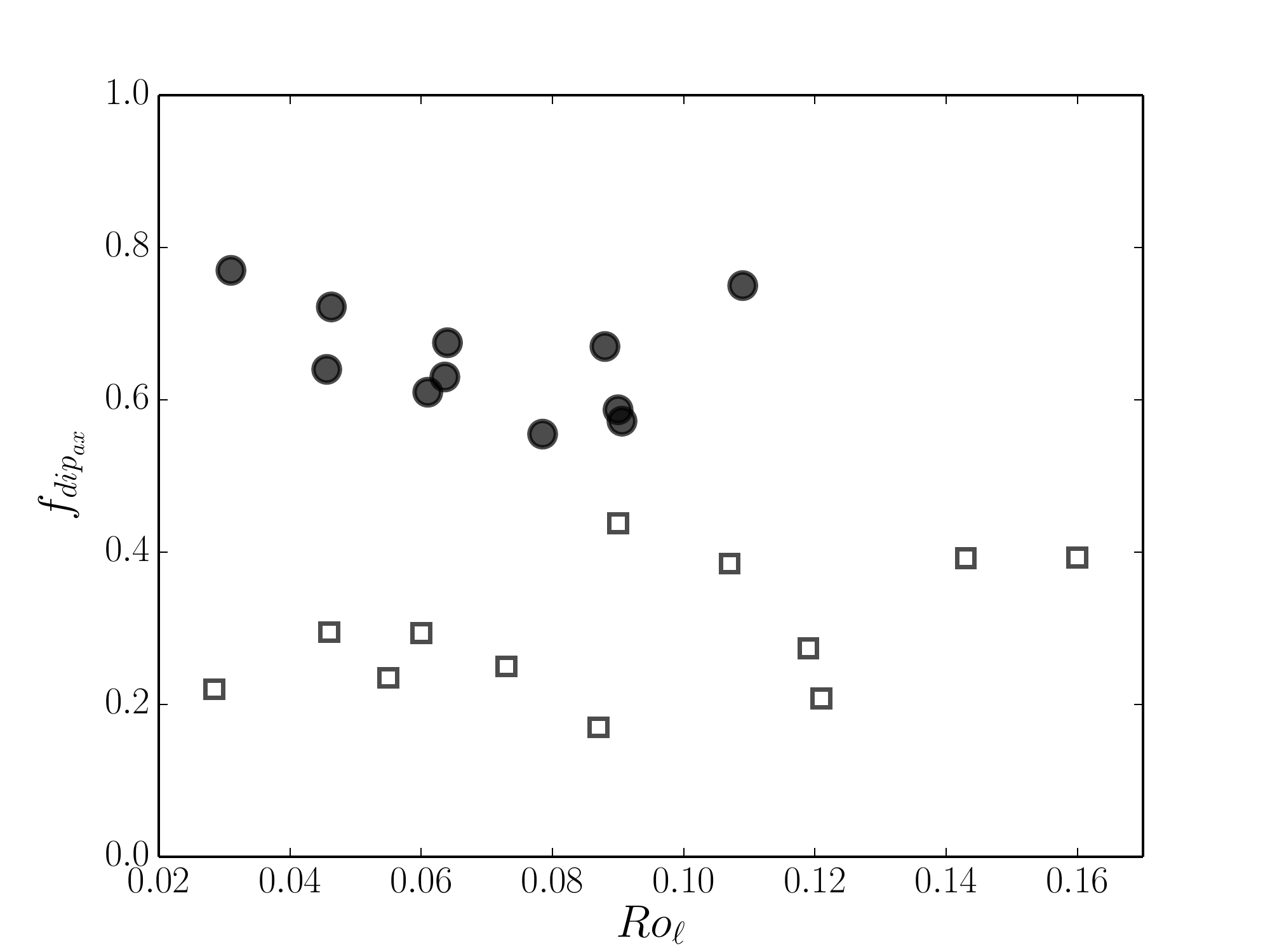}\label{sf:fdip20}}

  \caption{Left: dipolar (black circles) and multipolar (white
    squares) dynamos as a function of $\Ra/\Rac$ and \Pm{}, for
    $\Nrho=0.5$ (a), $\Nrho=1.5$ (c) and $\Nrho=2.0$ (e). A cross
    indicates the absence of a self-sustained dynamo. Right: the
    relative axial dipole field strength \fdipAX{} versus the local
    Rossby number for $\Nrho=0.5$ (b), $\Nrho=1.5$ (d) and $\Nrho=2.0$
    (f).  }\label{f:zones}
\end{figure*}

\section[]{Equations and methods}\label{s:eq}
Following \cite{jones11}, we rely on the LBR formulation of the
anelastic approximation \citep{lantz99,braginsky95}.  Actually, both
the model and the numerical methods used here are the same as in
\citet{schrinner2014} and \citet{raynaud2014} but we briefly recall
them for completeness. We consider a spherical shell of width~$d$ and
aspect ratio~$\aspectratio$, rotating about the $z$-axis at angular
velocity~$\vect{\Omega}$ and filled with a perfect, electrically
conducting gas with kinematic viscosity~$\nu$, thermal
diffusivity~$\kappa$, specific heat~$c_p$ and magnetic
diffusivity~$\eta$ (all assumed to be constant). Convection is driven
by an imposed entropy difference~$\Delta s$ between the inner and the
outer boundaries, and the gravity is given by
$\vect{g}=-GM\vect{\hat{r}}/r^2$, where $G$ is the gravitational
constant and $M$ the central mass.

The reference state is given by the polytropic equilibrium solution of
the anelastic system
\begin{equation}
\mP=P_c\,w^{n+1},\quad\mrho=\varrho_c\,w^n,\quad \mT=T_c\,w,\quad 
w=c_0+\frac{c_1 d}{r}, 
\label{ref_state}
\end{equation}
\begin{equation}
c_0=\frac{2w_0-\aspectratio-1}{1-\aspectratio},\quad
c_1=\frac{(1+\aspectratio)(1-w_o)}{(1-\aspectratio)^2},
\label{c0_c1}
\end{equation}
with 
\begin{equation}
w_0=\frac{\aspectratio+1}{\aspectratio\exp(N_\varrho/n)+1},\quad
w_i=\frac{1+\aspectratio-w_o}{\aspectratio}\,.
\label{w1_w0}
\end{equation}
In the above expressions, $n$ is the polytropic index and
$\Nrho=\ln{(\varrho_i/\varrho_o)}$ the number of density
scale-heights.  The values $P_c$, $\varrho_c$, and $T_c$ are the
reference-state density, pressure, and temperature mid-way between the
inner and outer boundaries, and serve as units for these variables.

Length is scaled by the shell width~$d$, time by the magnetic
diffusion time~$d^2/\eta$ and entropy by the imposed entropy
difference~$\Delta s$.  The magnetic field is measured in units of
$\sqrt{\Omega\varrho_c\mu\eta}$, where $\mu$ is the magnetic
permeability.  Then, the equations governing the system are
\begin{align}
  \begin{split}\label{mhd1}
    \frac{\partial\vect{v}}{\partial t} +
    \left(\vect{v}\cdot\nabla\right)\vect{v} &= Pm\,
    \bigg[-\frac{1}{\Ekman}\nabla\frac{P'}{w^n}
      +\frac{Pm}{Pr}Ra\frac{s}{r^2} \mathbf{\hat{r}}
      -\frac{2}{\Ekman}\,\mathbf{\hat{z}} \times\vect{v} \\ &\quad +
      \mathbf{F}_\nu+\frac{1}{\Ekman\,w^n}(\nabla\times\bb)\times\bb
      \bigg]\,,
  \end{split} \\
  \frac{\partial\bb}{\partial t} &= 
  \nabla\times(\vect{v}\times\bb)+\nabla^2\bb \,,\label{mhd2}\\
  \begin{split}\label{mhd3}
    \frac{\partial s}{\partial t}+\vect{v}\cdot\nabla s &= 
    w^{-n-1}\frac{Pm}{Pr}\nabla\cdot\left(w^{n+1}\,\nabla s\right) \\ 
    &\quad + \frac{Di}{w}\left[\Ekman^{-1}w^{-n}(\nabla\times \bb)^2+Q_\nu\right]\,,
  \end{split}\\
  \nabla\cdot \left(w^n\vect{v} \right)  &= 0\,,\label{mhd4}\\ 
  \nabla\cdot\bb &=  0\,.\label{mhd5}
\end{align}
The viscous force $\vect{F}_\nu$ in Eq.~\eqref{mhd1} is given by
\(\mathbf{F}_\nu=w^{-n}\nabla\mathbf{S}\), where $\vect{S}$ is the
rate of strain tensor
\begin{equation}
  S_{ij}=2w^n\left(e_{ij}-\frac{1}{3}\delta_{ij}\nabla\cdot \vect{v}\right),\quad
  e_{ij}=\frac{1}{2}\left(\frac{\partial v_i}{\partial x_j}+\frac{\partial
    v_j}{\partial x_i}\right) \, .
\end{equation}
Moreover, the expressions of the dissipation parameter~\(Di\) and the
viscous heating~\(Q_\nu\) in Eq.~\eqref{mhd3} are
\begin{equation}
Di=\frac{c_1Pr}{Pm Ra} \, ,
\end{equation} 
and
\begin{equation}
Q_\nu=2\left[e_{ij}e_{ij}-\frac{1}{3}(\nabla\cdot\vect{v})^2\right] \, .
\end{equation}

We impose stress-free boundary conditions for the velocity field at
both the inner and the outer spheres, the magnetic field matches a
potential field inside and outside the fluid shell, and the entropy is
fixed at the inner and outer boundaries. Besides, both weak and strong
field initial conditions have been tested for all models, since the
system may exhibit hysteretic transitions between dynamo branches when
stress-free boundary conditions are used.

\begin{figure*}
  \centering
  \subfigure[]{
  \includegraphics[width=\mafig]{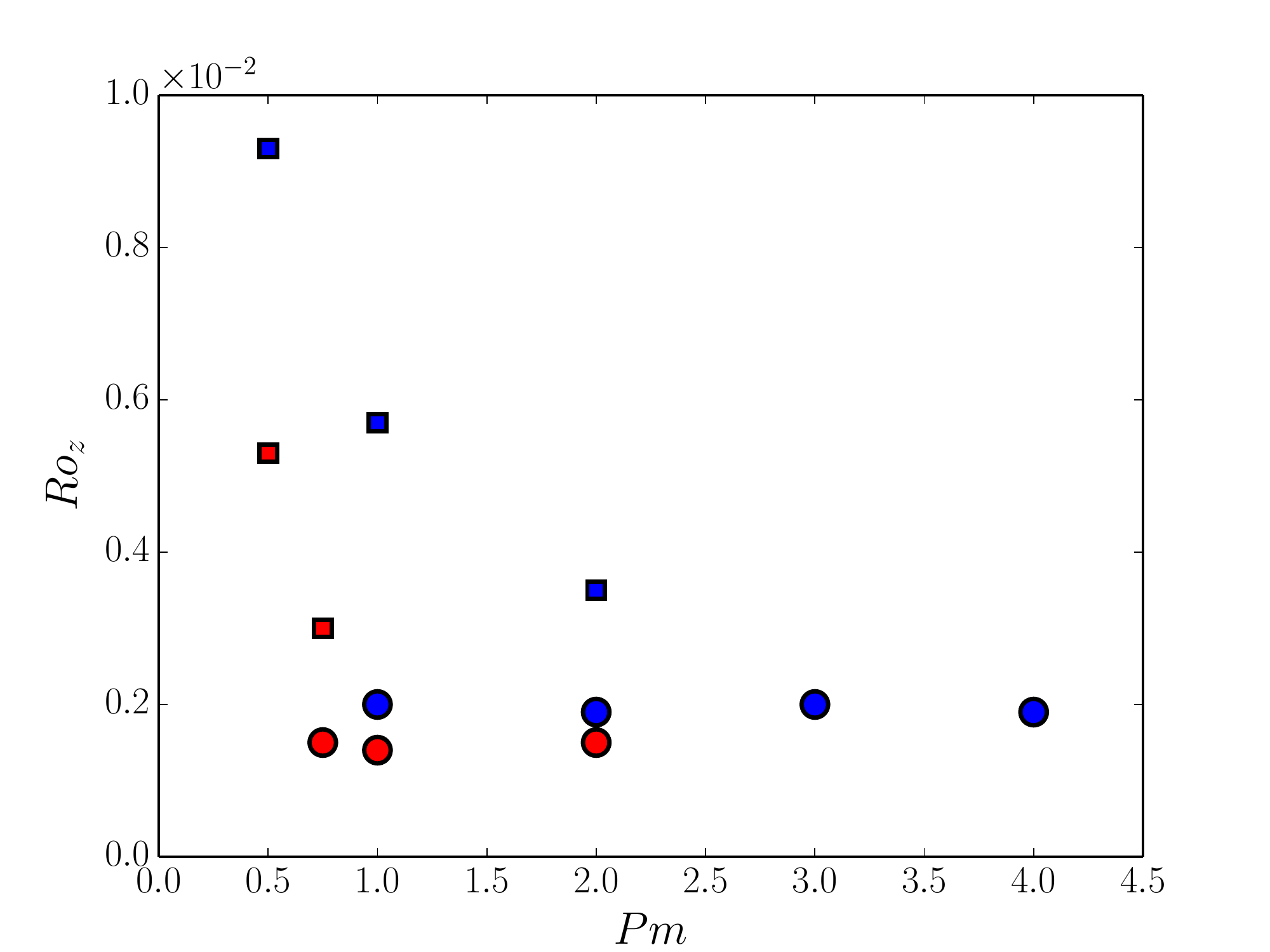}\label{f:roz}
  }
  \subfigure[]{
  \includegraphics[width=\mafig]{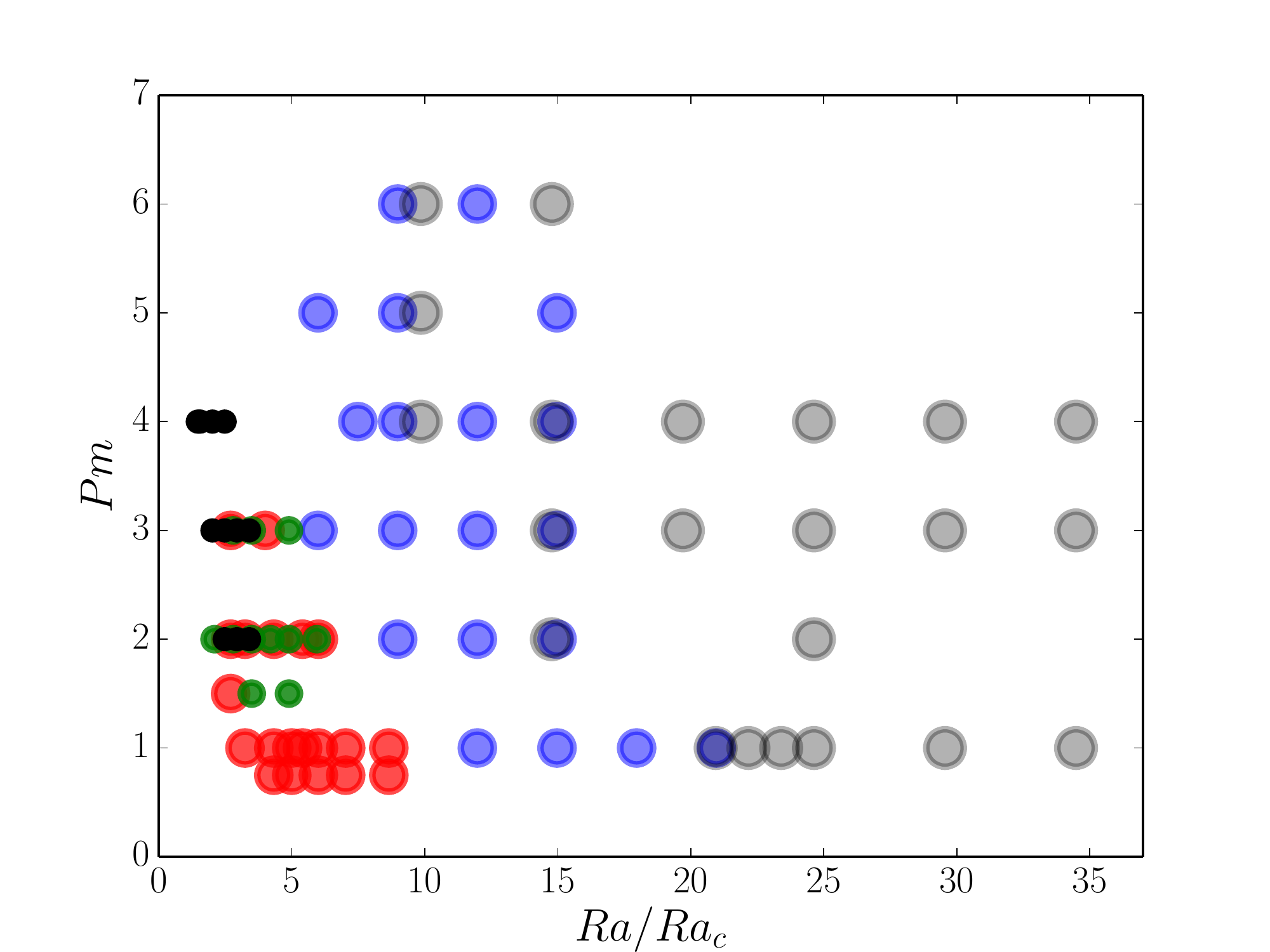}\label{f:dip}}

  \caption{(a): Evolution of the zonal Rossby number as a function of
    \Pm{} for a dynamo models with $\Ra = 4\times10^6$, at $\Nrho=0.5$
    (blue) and $\Nrho=1.5$ (red). Circles (squares) stands for dipolar
    (multipolar) dynamos. (b): Dipolar dynamos in the parameter space
    ($\Ra/\Rac$, \Pm), for increasing density stratifications:
    $\Nrho=0.1$ (grey), $\Nrho=0.5$ (blue) $\Nrho=1.5$ (red),
    $\Nrho=2.0$ (green) and $\Nrho=2.5$ (black).}
\end{figure*}
\begin{figure*}
  \centering
  \subfigure[]{
  \includegraphics[width=\mafig]{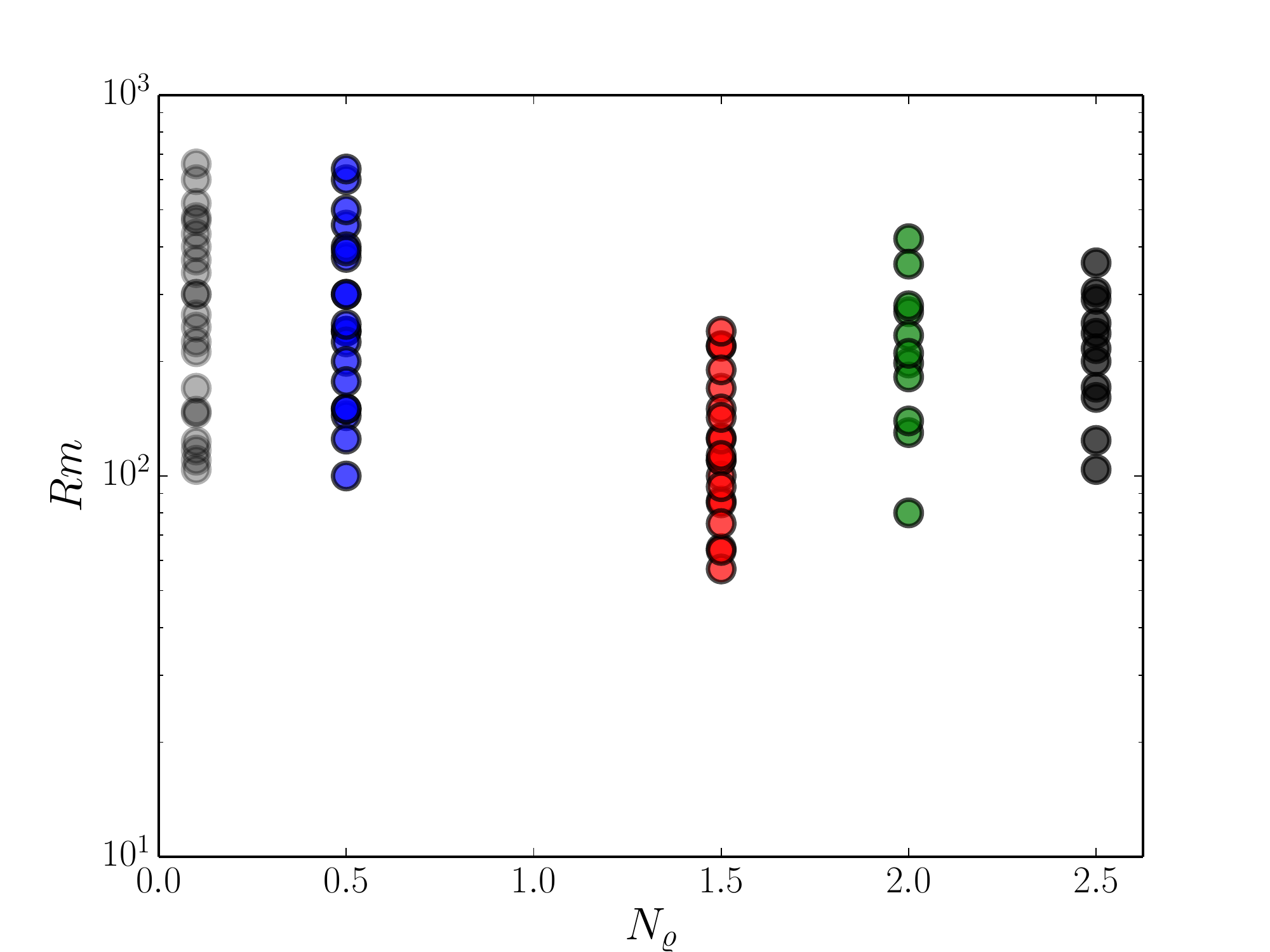}\label{f:rmc}
  }
  \subfigure[]{
  \includegraphics[width=\mafig]{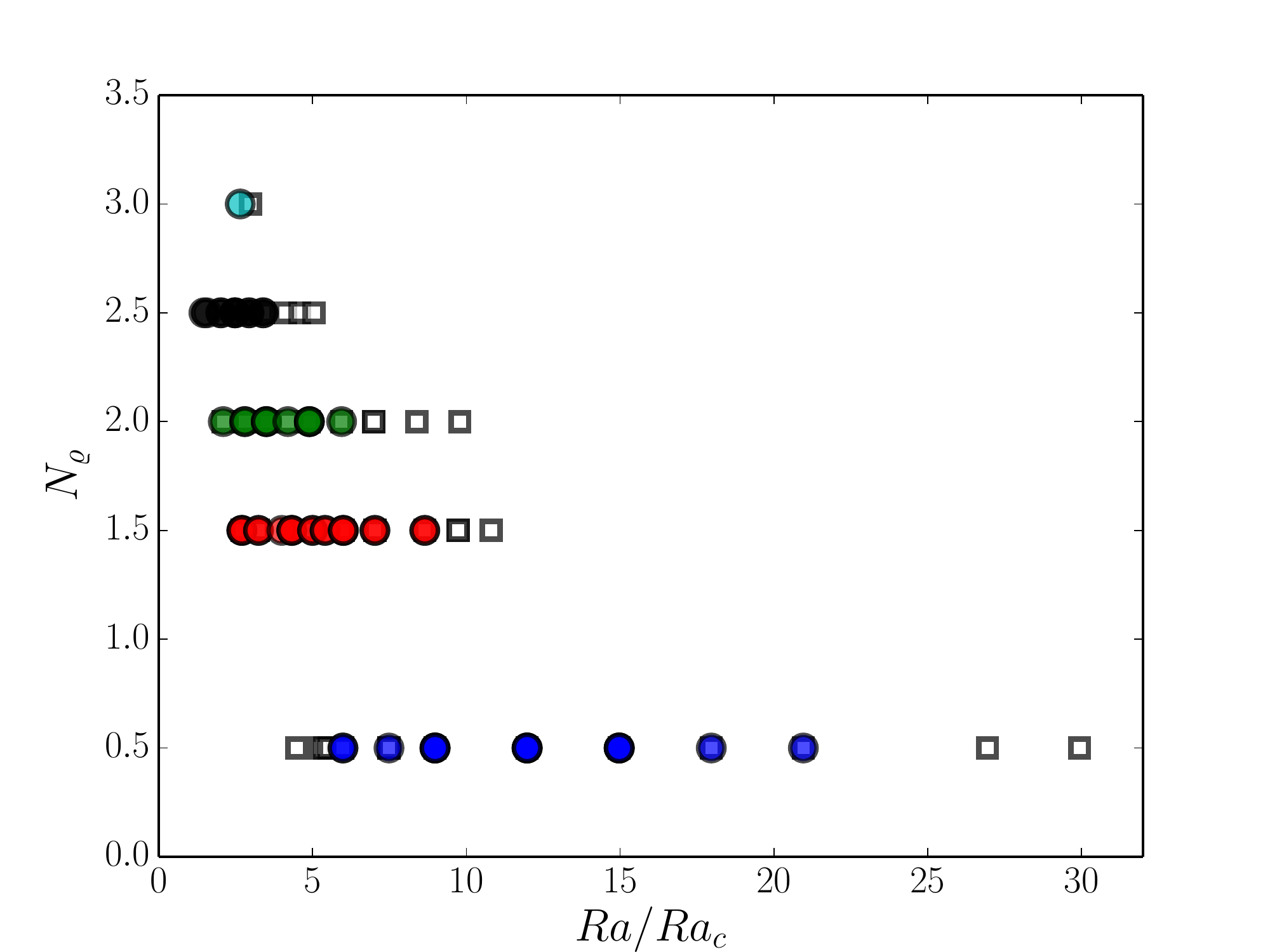}\label{f:np}
  }
  \caption{(a): Magnetic Reynolds number as a function of \Nrho{} for
    dipolar dynamos. (b): Our sample of dipolar (circles) and
    multipolar (squares) dynamos in the parameter space ($\Ra/\Rac$,
    \Nrho).}\label{f:3}
\end{figure*}

The system of equations (\ref{mhd1})--(\ref{mhd5}) involves seven
control parameters, namely the Rayleigh number~$\Ra=GMd\Delta s /
(\nu\kappa c_p)$, the Ekman number~$\Ekman =\nu / (\Omega d^2) $, the
Prandtl number~$\Prandtl = \nu / \kappa $, and the magnetic Prandtl
number~$\Pm = \nu / \eta$, together with the aspect
ratio~\aspectratio{}, the polytropic index~$n$, and the number of
density scale-heights~\Nrho{} that define the reference state.  We
choose $\Ekman=10^{-4}$, $\Prandtl=1$ and $n=2$.  Different from
\cite{gastine12}, we also kept the central gravity profile and the
aspect ratio $\aspectratio=0.35$ fixed for all simulations, but varied
the magnetic Prandtl number, which turns out to be a key point to
understand the partial divergence of our conclusions.

The equations are integrated in average for one magnetic diffusion
time with the anelastic version of \textsc{parody}
\citep{dormy98,schrinner2014}.\footnote{The integration times range
  from 0.63 to 5.2 magnetic diffusion times (for the models
  \texttt{101m} and \texttt{004m}, respectively).}  The vector fields
are transformed into scalars using the poloidal--toroidal
decomposition. The equations are then discretized in the radial
direction with a finite-difference scheme; on each concentric sphere,
variables are expanded using a spherical harmonic basis. The
coefficients of the expansion are identified with their degree~$\ell$
and order~$m$. Typical resolutions are 288 points in the radial
direction (up to 320 points).  The spectral decomposition is truncated
at a hundred modes (up to $\ell_\text{max} \sim m_\text{max} \le
128$), in order to observe for both spectra a decrease of more than
two orders of magnitude over the range of $l$ and $m$.  The highest
resolutions are required for the models with the highest density
stratification ($\Nrho=3$).

The amplitudes of the velocity and the magnetic fields are measured in
terms of the Rossby number $\Ro=\sqrt{2 \Ek} \Ekman/\Pm$ and Lorentz
number $\Lo=\sqrt{2 \Em}\Ekman/\Pm$, where \Ek{} and \Em{} are the
energy densities integrated over the fluid shell,
\begin{equation}
\Ek = \frac{1}{2\, V}\int_V\,w^n\bv^2 \dif v 
\quad \text{and} \quad
\Em = \frac{1}{2\, V}\frac{Pm}{E}\int_V\,\bb^2 \dif v
\,.
\label{emag}
\end{equation}
Likewise, the measure of the mean zonal flow is given by the zonal
Rossby number \Roz{} based on the averaged toroidal axisymmetric
kinetic energy. 

We also define a local Rossby number $\Rol = Ro_c\,\lc/\pi$ based on
the mean harmonic degree~\lc{} of the velocity component $\vect{v}_c$
from which the mean zonal flow has been subtracted
\citep{schrinner12,schrinner2014},
\begin{equation}
  \lc = \sum_\ell \ell\frac{\langle w^n\,
    (\vect{v}_c)_\ell\cdot(\vect{v}_c)_\ell\rangle} {\langle w^n \,
    \vect{v}_c\cdot\vect{v}_c \rangle} \,,
\end{equation}
where the brackets denote an average over time and radii. The
contribution of the mean zonal flow is removed for calculating
$\Ro_c$.

Furthermore, as the stratification is increased, it turns out that it
is useful to examine the variations with depth of the local Rossby
number, defined in such a way that $\Rol = \int_{\ri}^{\ro} \Rol(r)
r^2 \dif r$. We found that it is more suitable to slightly adapt our
initial definition and investigate the radial dependence of
$\Rolr(r)$, which differs from $\Rol(r)$ in so far as the velocity is
not weighted by the reference density profile~$w^n$. We also checked
that, in our range of \Nrho{}, both estimates of a characteristic
velocity do not make a qualitative difference on the volume-averaged
quantities.  For instance, the difference between the values of the
magnetic Reynolds number $\Rm = \Ro \Pm /\Ekman$ is about 1 per cent
at $\Nrho=0.5$.  Of course, it increases with \Nrho{}: energy-based
estimates lead to lower values about 7 and 10 per cent for $\Nrho=2.0$
and $\Nrho=2.5$, respectively. However, this does not change our
conclusions, and that is why we do not adapt our definition for
volume-averaged quantities.

\section[]{Results}\label{s:results}

\subsection{Bistability}

Figure~\ref{f:zones} shows the distribution of dipolar and multipolar
dynamos in the parameter space ($\Ra/\Rac$, \Pm) (left-hand panels),
together with the corresponding dipolarities (right-hand panels), for
increasing density stratifications from top to bottom. One can see
that several examples of bistable pairs are displayed.  Bistability
is commonly known for Boussinesq and anelastic models, and is related
to the use of stress-free boundary conditions that allows for the
growth of stronger zonal winds
\citep{sasaki2011,schrinner12,gastine12}.  For $\Nrho=0.5$, the regime
diagram in Fig.~\ref{sf:05} does not qualitatively differ from what we
can observe in the Boussinesq regime. As we found in
\cite{raynaud2014}, the multipolar branch undergoes a supercritical
bifurcation as \Ra{} is increased, whereas the dipolar one still loses
its stability in favour of the multipolar branch at low Rayleigh and
magnetic Prandtl numbers.  When increasing the density contrast
\Nrho{} to 1.5, one can note in Fig.~\ref{sf:15} that the overlap
between the two branches shrinks.  At $\Nrho=2$, we do not observe a
bistable case. More generally, for all models of our sample with a
density stratification $\Nrho \ge2$, the saturated field of the dynamo
is not anymore sensitive to the amplitude of the initial magnetic
field.  To understand this evolution from Boussinesq models to
anelastic models with moderate stratification ($\Nrho\le1.5$), it is
worth stressing that the transition process from the multipolar to the
dipolar branch triggered by the increase of \Pm{} still applies to our
sample of models \citep[see][]{schrinner12}.  Figure~\ref{f:roz}
illustrates on a few cases the progressive merging of the multipolar
branch which is indeed lost when its zonal Rossby number becomes
comparable to the zonal Rossby number of the dipolar branch.  For a
given Rayleigh number, the fact that the mean zonal flow of the
multipolar branch decreases with \Pm{} (and eventually becomes too
small to prevent the growth of the axial dipole) is actually the
limiting factor of the upper extent of the multipolar branch in the
left-hand panels of Fig.~\ref{f:zones}. This also emphasizes the
essential role played by differential rotation in the dynamo mechanism
of the multipolar branch, often accounted for in terms of
$\Omega$-effect.

Interestingly, the zonal Rossby number for multipolar dynamos
substantially decreases between $\Nrho=0.5$ and 1.5 (see the blue and
red squares in Fig.~\ref{f:roz}), while it remains of the same order
for dipolar dynamos. Hence, the available range of \Pm{} for the
multipolar solution is reduced, which therefore explains the relative
shrinking of the bistable region when comparing Figs~\ref{sf:05} and
\ref{sf:15}. The simplest argument to understand this downtrend is
given by the comparison of the $x$-axis in Fig.~\ref{f:zones}, which
reveals that the dynamo onset moves closer to the onset of convection
when the density stratification is increased, as mentioned by
\cite{gastine12}. Indeed, despite changing the value of \Nrho{}, we
found that the Rayleigh numbers we had to consider always stay of the
order of $10^6$.  At the same time, the critical Rayleigh number for
the linear onset of convection monotonically increases with
\Nrho{}. From table B.1 in \cite{schrinner2014}, we have in our case
the following values of $3.34\times10^5$, $9.25\times 10^5$ and
finally $1.43\times10^6$ for the sequence of density stratifications
$\Nrho=0.5$, 1.5 and 2.0, respectively.

\subsection{Dipole onset}

The density stratification strongly impacts on the stability domain of
the dipolar branch, as we clearly see in Fig.~\ref{f:dip}. In this
figure, we included data from \cite{raynaud2014} in order to better
highlight the differences with Boussinesq simulations.  For moderate
values of \Nrho{} at a fixed \Pm{}, the critical value of $\Ra/\Rac$
at which it is possible to sustain a dipolar dynamo rapidly falls off
(up to a factor of 4 if we consider the line $\Pm=1$). However, this
tendency hardly persists once we reach $\Ra/\Rac \sim 5 $ for
$\Nrho=1.5$, and the further increase of \Nrho{} mainly affects the
critical magnetic Prandtl number \Pmc{} below which it is not possible
to sustain a dipolar dynamo. In our sample of models, the increase of
\Pmc{} becomes effective for $\Nrho{} \geq 2$, but we already reported
it as a general tendency in \cite{schrinner2014}.  Figure~\ref{sf:15}
enables us to conclude that $0.5 <\Pmc\le 0.75$ for $\Nrho=1.5$,
whereas from Fig.~\ref{sf:20}, it is clear that $\Pmc > 1$ for
$\Nrho=2$.

The fact that dipolar dynamos are found closer to the convection
threshold as \Nrho{} increases can be more or less readily understood
if one notices that, despite the increase of the density
stratification, the critical magnetic Reynolds number \Rmc{} of the
dipolar branch does not significantly vary, but stays in first
approximation of the order of $10^2$, as shown in Fig.~\ref{f:rmc}.
Then, if we take this as a necessary condition to obtain a dipolar
solution, and given the fact that for a constant value of $\Ra/\Rac$
the flow amplitude increases with \Nrho{} \citep{gastine12}, it
explains why the dipolar branch can be found closer to the onset of
convection when the stratification increases.  However, we will see in
the next subsection that, as \Nrho{} is further increased, not only
does the dipolar branch occur closer to the onset of convection, but
also higher magnetic Prandtl numbers have to be considered to maintain
a sufficiently high \Rm{} while preventing the collapse of the dipole.

\subsection{Dipole collapse}

\begin{figure}
  \centering
  \includegraphics[width=\mafig]{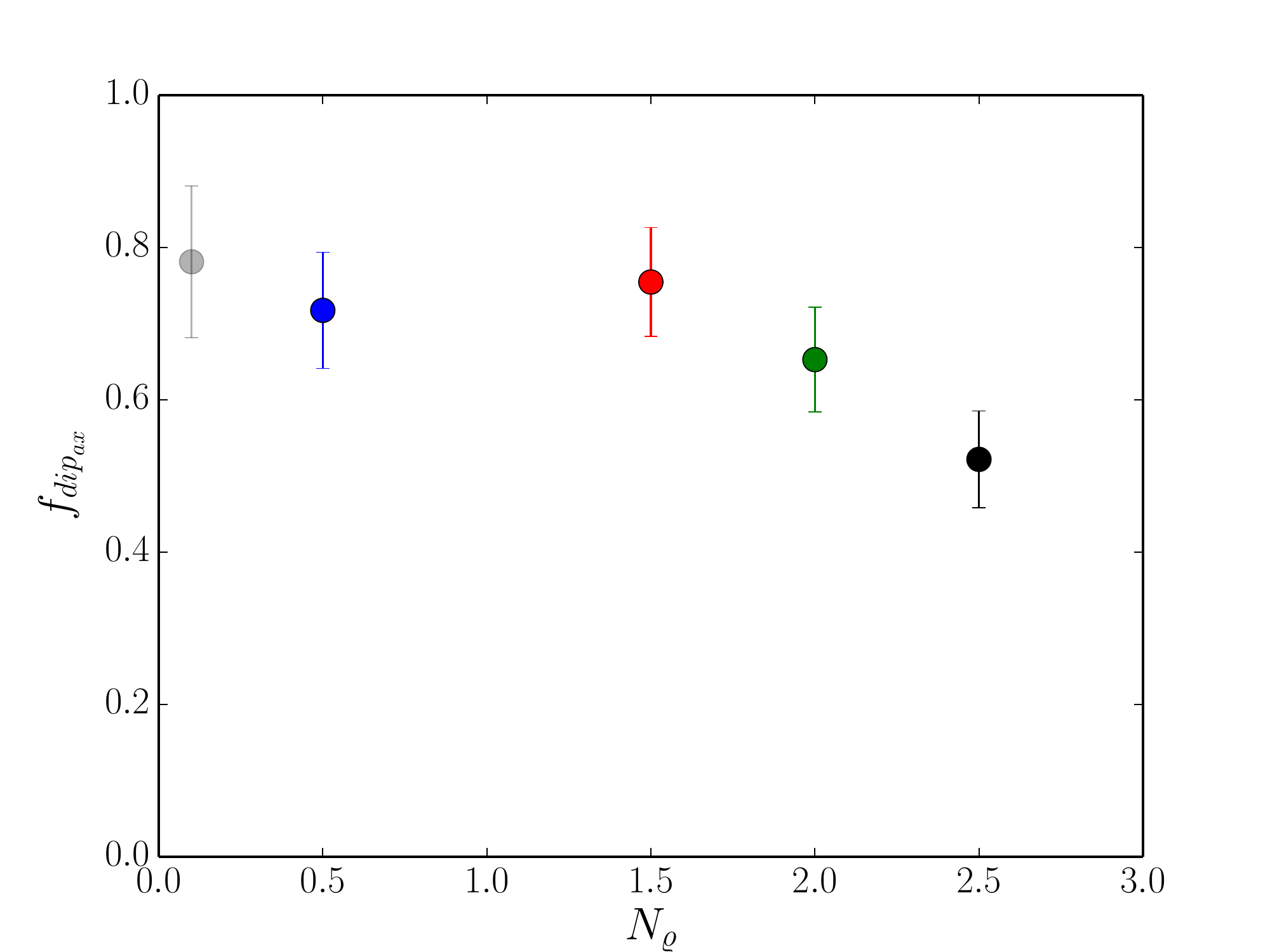}
  \caption{Average values of \fdipAX{} for dipolar dynamos as a
    function of \Nrho{}. Error bars represent the standard
    deviation. The average is done with 11 models for
    $\Nrho=2.5$.}\label{f:fdip_rol}
\end{figure}

\begin{figure*}
  \centering
  \subfigure[$\Nrho=1.5$]{
    \includegraphics[width=\mafig,trim=50 20 40 20,clip=true]{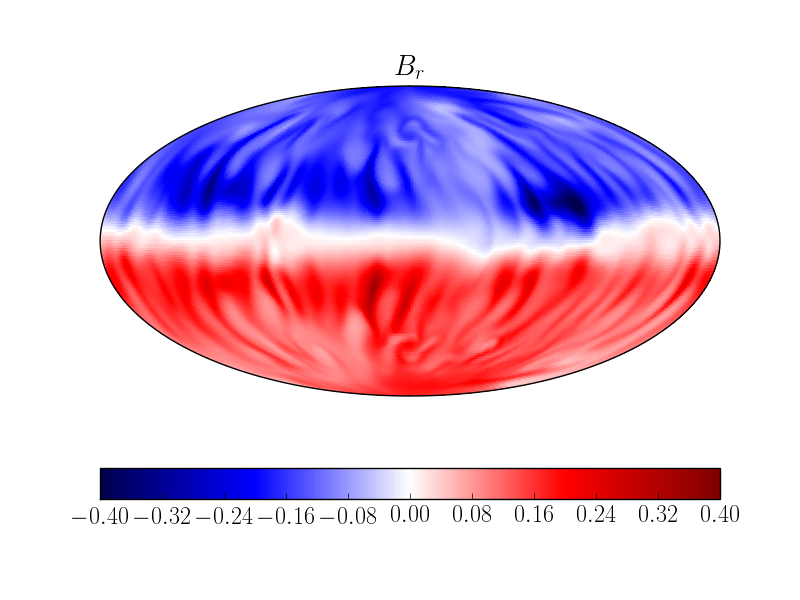}
  \label{sf:br15}}
  \subfigure[$\Nrho=1.5$]{
    \includegraphics[width=\mafig,trim=50 20 40 20,clip=true]{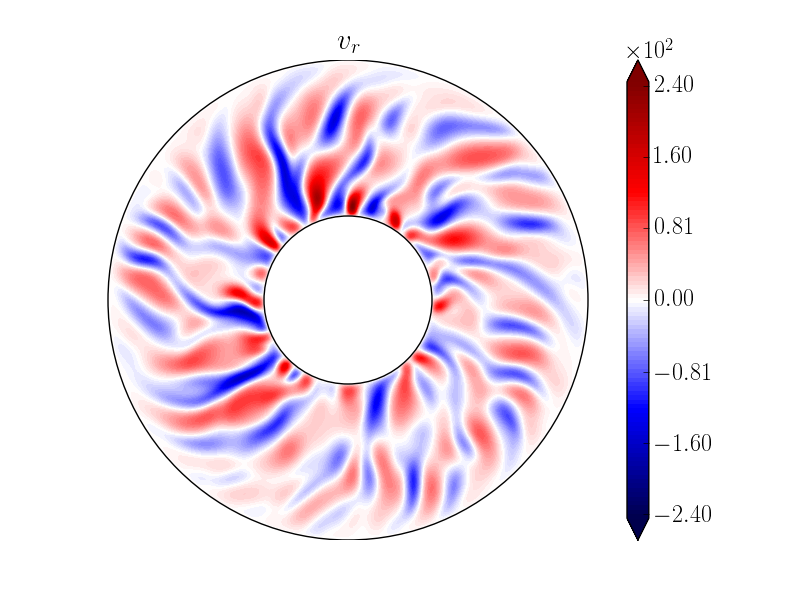}
  \label{sf:vr15}}
  \caption{Snapshot of $B_r\left(r=\ro\right)$ (a) and equatorial cut
    of $v_r$ (b) for a dipolar dynamo with $\Nrho=1.5$, $\Pm=0.75$,
    $\Ra=4.625\times10^{6}=5 \Rac$.}\label{f:snap15}
\end{figure*}

\begin{figure*}
  \subfigure[$\Nrho=2.5$]{
    \includegraphics[width=\mafig,trim=50 20 40 20,clip=true]{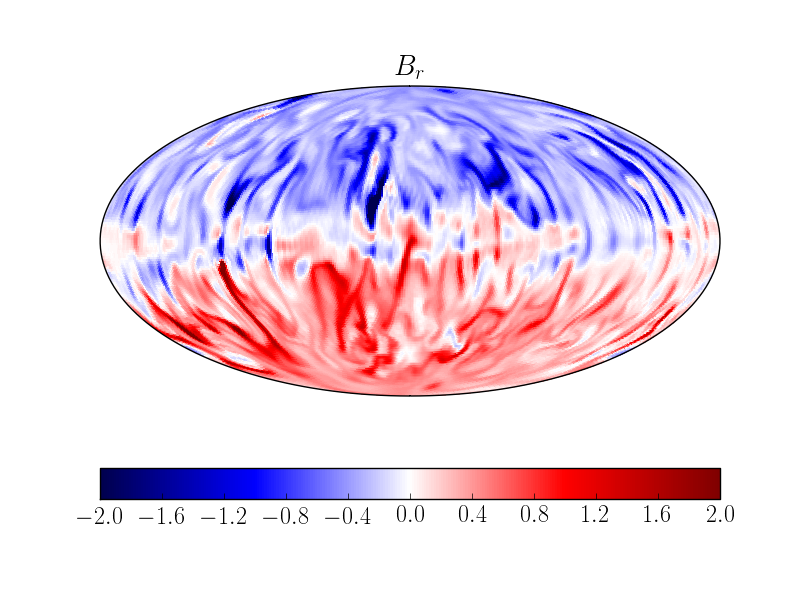}
  \label{sf:br25}}
  \subfigure[$\Nrho=2.5$]{
    \includegraphics[width=\mafig,trim=50 20 40 20,clip=true]{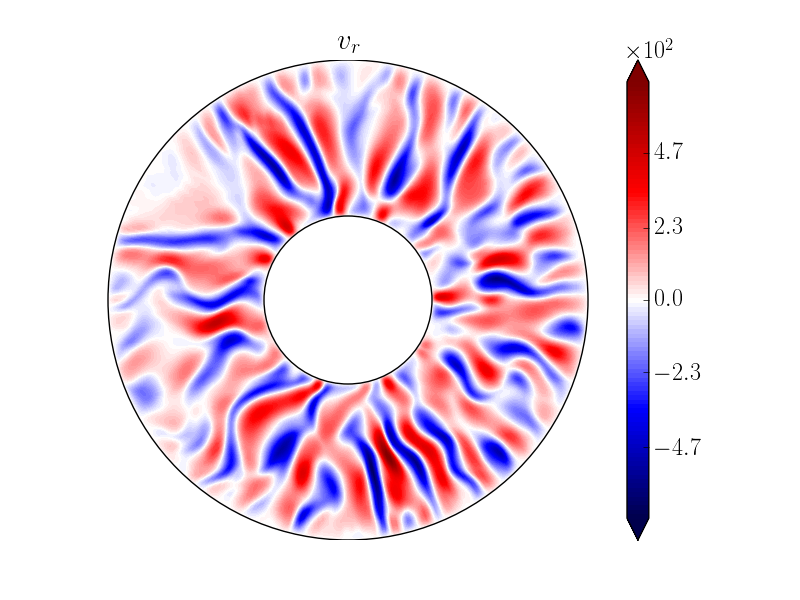}
  \label{sf:vr25}}

  \caption{Snapshot of $B_r\left(r=\ro\right)$ (a) and equatorial cut
    of $v_r$ (b) for a dipolar dynamo with $\Nrho=2.5$, $\Pm=2$,
    $\Ra=7.40\times10^{6}=3.4\Rac$.}\label{f:snap25}
\end{figure*}

Another striking feature that arises when investigating the stability
domain of the dipolar branch is that the range of Rayleigh numbers
over which it extends becomes smaller and smaller as \Nrho{} is
increased.  This is clearly visible in Fig.~\ref{f:np} that shows for
different \Nrho{} the transition from the dipolar to the multipolar
branch resulting from the increase of $\Ra$.  In other words, at this
moderate value of the Ekman number, dipolar dynamos are confined in a
narrower and narrower window of Rayleigh numbers, which explains why
dipolar solutions may seem more difficult to obtain at higher \Nrho{},
despite comparable critical magnetic Reynolds numbers. As for the
modification of the dynamo onset, this can be related to the fact that
for a given value of $\Ra/\Rac$, the Rossby number \Ro{} increases
with \Nrho{}.\footnote{For instance, one can compare the models
  \texttt{008d}, \texttt{055d} and \texttt{083d} for which $\Ra/\Rac
  \sim 6$ and an increasing \Ro{} of $4.8 \times 10^{-3}$, $1.3 \times
  10^{-2}$ and $1.8 \times 10^{-2}$, respectively, or else the models
  \texttt{021d}, \texttt{051d} and \texttt{096d} that have a similar
  Rossby number of 0.01, but for which $\Ra/\Rac$ is about 12, 5 and
  2.9, respectively.} The transition from a dipolar to a multipolar
solution triggered by an increase of \Ra{} is related to the fact that
inertia becomes significant in the force balance. We know from
\citet{christensen06} that this transition can be measured by a local
Rossby number \Rol{} based on a characteristic length-scale of the
flow. We find that the collapse of the dipole still occurs for $\Rol
\sim 0.1$ when $\Nrho\leq 2$, which is consistent with the results in
\citet{gastine12}.  This is illustrated by Figures~\ref{sf:fdip05},
\ref{sf:fdip15} and \ref{sf:fdip20} which show the relative axial
dipole field strength \fdipAX{} computed at the outer sphere, as a
function of \Rol{}.  In Fig.~\ref{sf:fdip05}, the very low values of
\fdipAX{} at low \Rol{} are characteristics of multipolar dynamos
dominated by an equatorial dipole component. We showed in
\cite{raynaud2014} that this magnetic configuration arises close to
the dynamo onset and when convective cells are localized close to the
inner sphere. However, we know from hydrodynamic studies that the
convection cells move towards the outer shell when the stratification
is increased \citep{jones_lin,gastine12a}, which explains why this
feature tends to disappear in Figs~\ref{sf:fdip15} and
\ref{sf:fdip20}.  Besides, we see in Fig.~\ref{f:fdip_rol} that the
values of \fdipAX{}
tend to decrease with \Nrho{}, which is also clear if we focus for
instance on the dipolar branch in Fig.~\ref{sf:fdip20} for which
$\fdipAX{}<0.8 $.  As expected, this indicates that the small magnetic
scales at the outer surface are favoured with the increase of the
stratification. This is also clearly confirmed by the comparison of
the radial magnetic fields at the outer surface of the model, as shown
in the left-hand panelsleft
 of Figs~\ref{f:snap15} and \ref{f:snap25}.
Finally, we also report the existence of multipolar dynamos whose
dipolarity displays strong variations in time. This leads to averaged
values of $\fdipAX \sim 0.5$, as one can notice in
Fig.~\ref{sf:fdip15}. These dynamos usually exhibit a relatively
strong axial dipole component which undergoes reversals during which
the value of \fdipAX{} decreases drastically. \cite{duarte2013} also
reported similar behaviour for dynamo models with a variable
electrical conductivity.
\begin{figure*}
  \centering
  \subfigure[]{
    \includegraphics[width=0.32\textwidth]{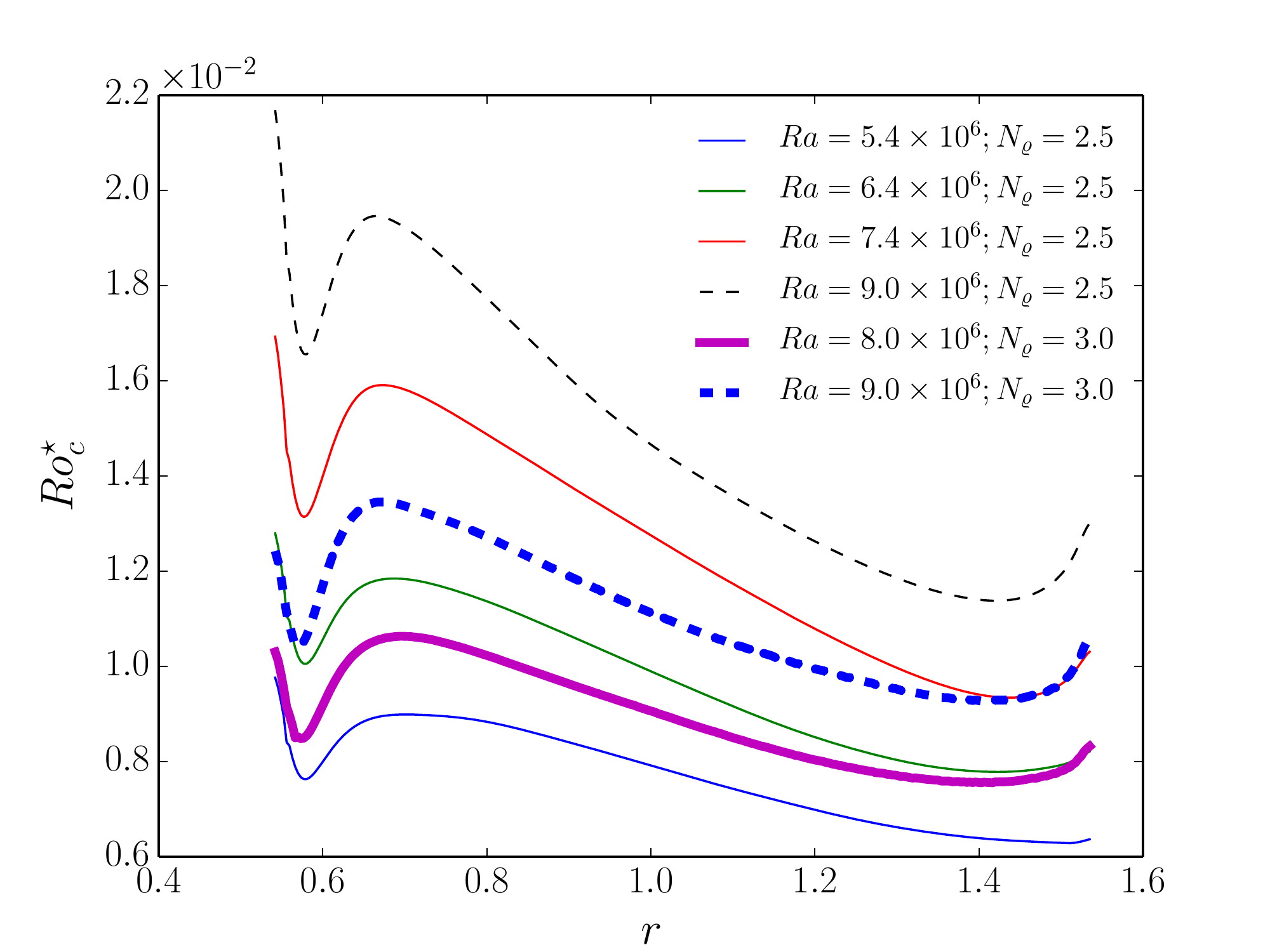}\label{sf:rocstar}}
  \subfigure[]{
    \includegraphics[width=0.32\textwidth]{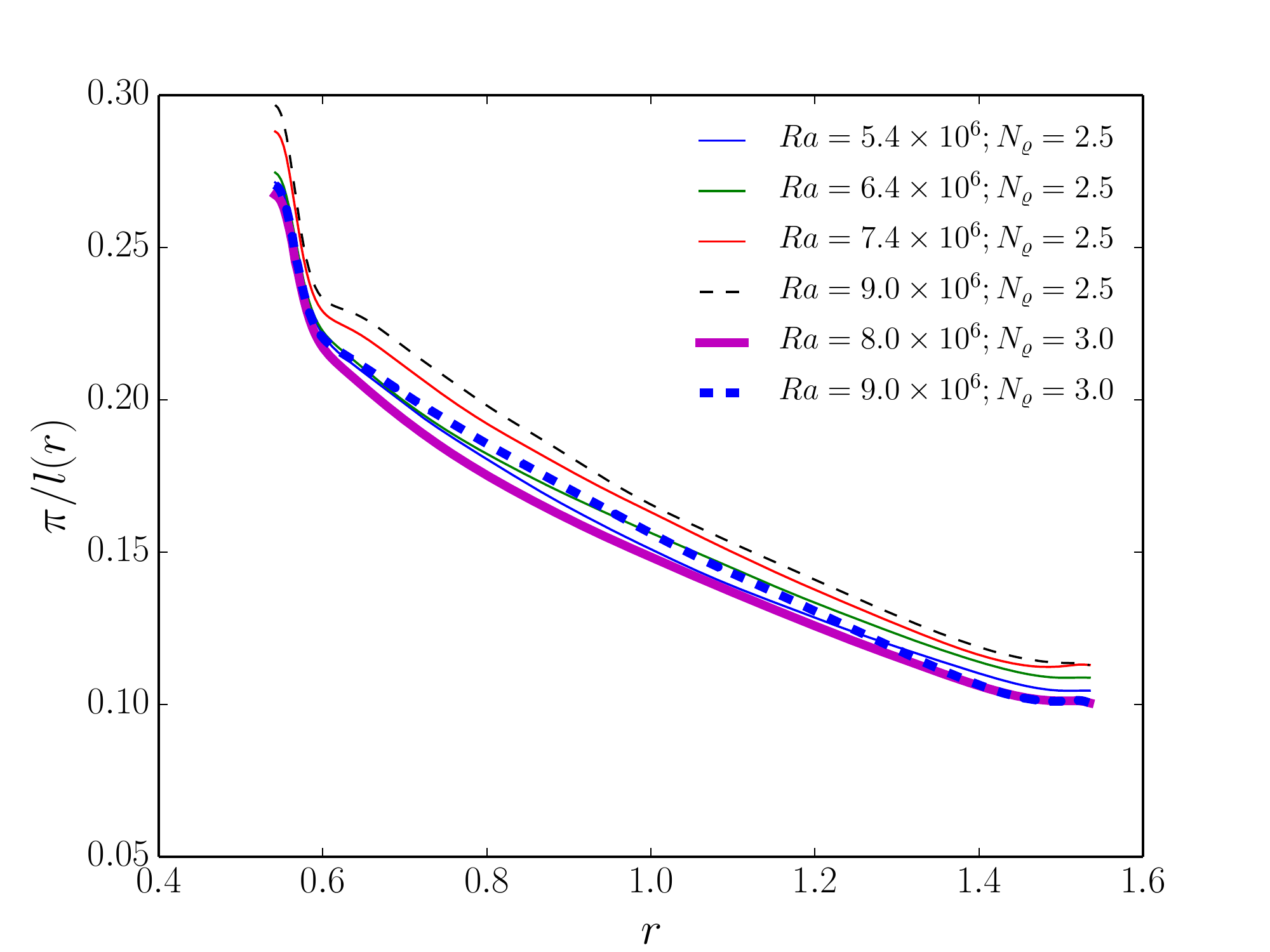}\label{sf:rl}}
  \subfigure[]{
    \includegraphics[width=0.32\textwidth]{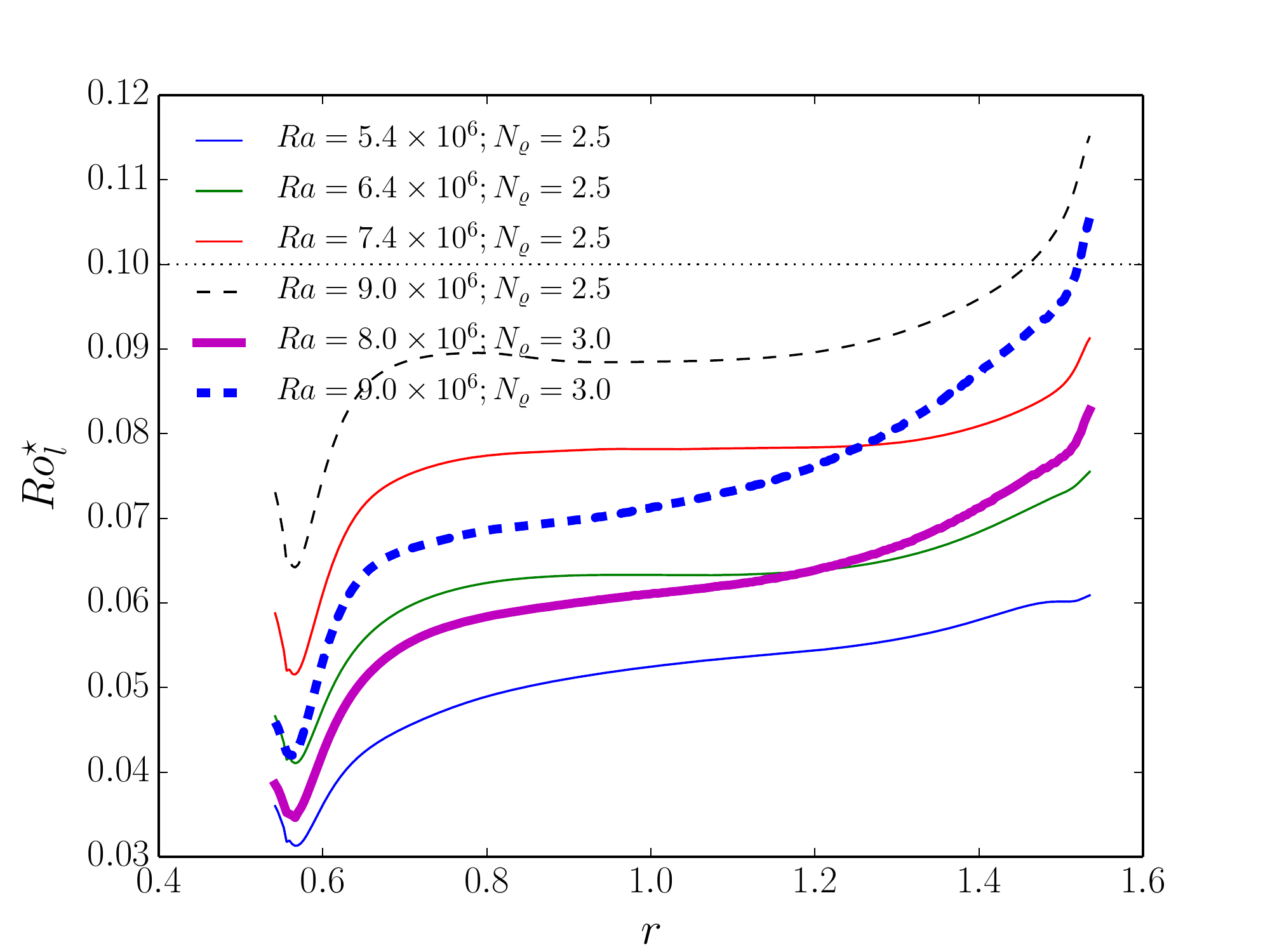}\label{sf:rrol}}
  \caption{The convective Rossby number (a), the convective
    length-scale (b) and the local Rossby number (c) as a function of
    radius for dipolar (solid lines) and multipolar (dashed lines)
    dynamos at ($\Nrho=2.5$, $\Pm=2$) (thin lines) and ($\Nrho=3$,
    $\Pm=4$) (thick lines).}\label{f:rolr}
\end{figure*}

For $\Nrho > 2.0$, we found that the dipole collapse tends to occur at
values of \Rol{} lower than $0.1$. However, it is likely that a
volume-averaged quantity becomes less relevant when applied to models
with a substantial stratification. For instance, we see in
Figs~\ref{sf:vr15} and \ref{sf:vr25} that the smaller structures that
develop at $\Nrho=2.5$ are confined close to the outer boundary,
whereas there are no significant differences in the radial flow at
mid-depth.  Thus, we also examined the radial dependence of the
different components of the local Rossby number $\Ro_l^\star$, which
is computed as the product of two terms: a convective Rossby number
based on the velocity field~$\vect{v}_c$ from which the mean zonal
flow has been subtracted (see Fig.~\ref{sf:rocstar}) and a
characteristic length-scale based on the mean harmonic degree of
$\vect{v}_c$ (see Fig.~\ref{sf:rl}). We find that the monotonicity of
$\Ro_l^\star$ changes as \Nrho{} is increased.  Indeed, for low
stratifications, $\Ro_l^\star(r)$ mainly decreases with radius,
whereas for $\Nrho\ge 2.5$ it becomes an increasing function of $r$
that steepens slightly close to the outer
surface. Figure~\ref{sf:rrol} shows the evolution of $\Ro_l^\star(r)$
for increasing Rayleigh numbers up to the loss of the dipolar
solution, at $\Nrho=2.5$ and 3.0. When the transition to the
multipolar branch is reached, we see that \Rolr{} tends to increase
faster close to the outer surface, while the volume-averaged value can
stay below the critical value of $0.1$.  Thus, it seems that inertia
still causes the collapse of the dipolar branch, despite the fact that
the usual local Rossby number criterion is not appropriate to separate
the two dynamo branches for significant density stratifications.

\subsection{Dynamo mechanisms}

\begin{figure*}
  \centering
  \subfigure[$\Nrho=1.5$]{
    \includegraphics[width=\mafigME,trim=150 0 100
    0,clip=true]{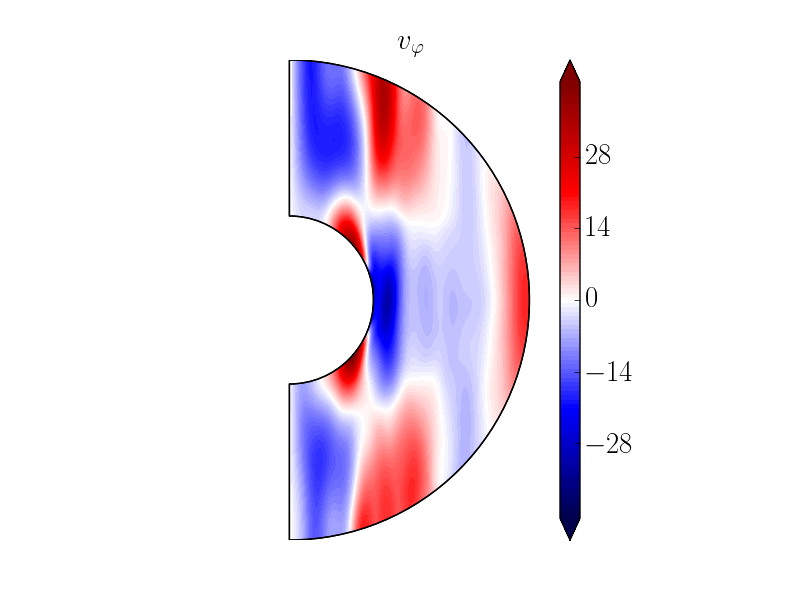}\label{sf:vp15}
  }
  \subfigure[$\Nrho=1.5$]{
    \includegraphics[width=\mafigME,trim=150 0 100
    0,clip=true]{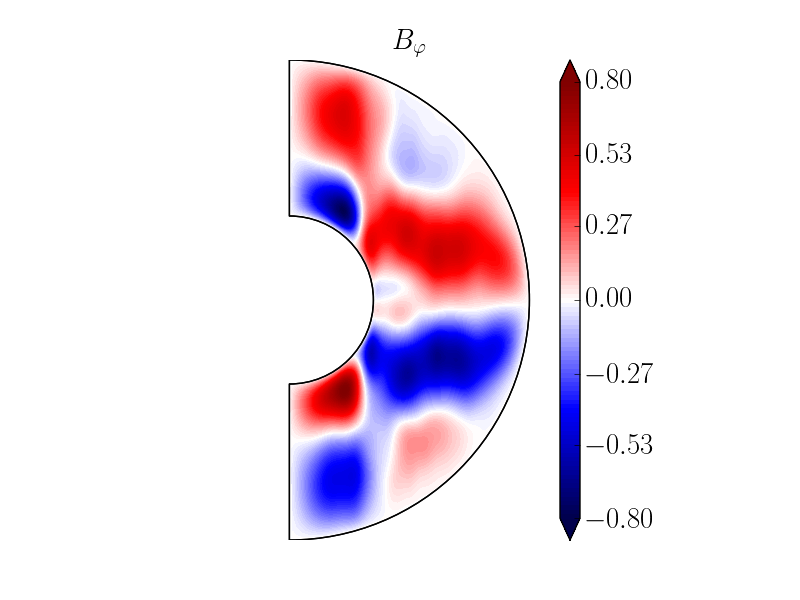}\label{sf:bp15}
  }
  \subfigure[$\Nrho=1.5$]{
    \includegraphics[width=\mafigME,trim=150 0 100
      0,clip=true]{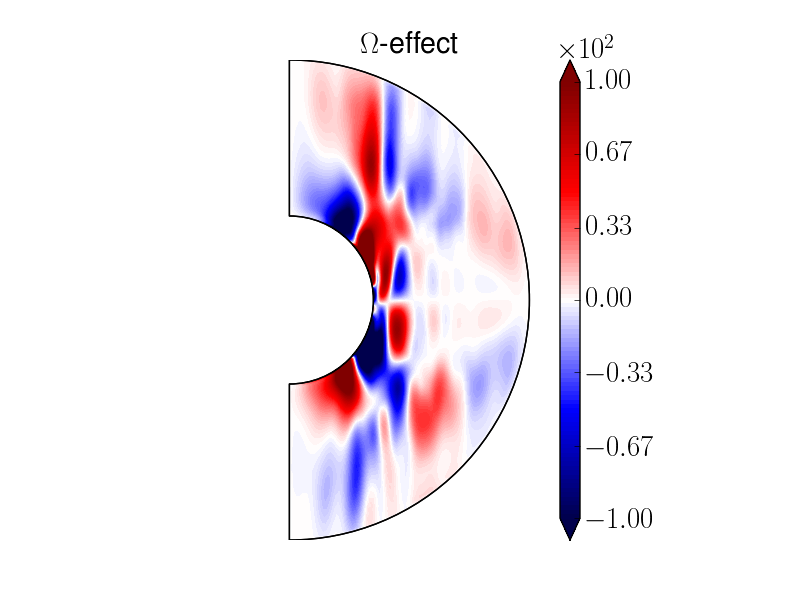}\label{sf:om15}
    }
  \caption{Time-averaged axisymmetric component of the azimuthal
    magnetic field (b) and velocity field (a) for a dipolar
    dynamo with $\Nrho=1.5$, $\Ra=4.625\times10^{6}$,
    $\Pm=0.75$.}\label{f:npa15}
\end{figure*}
\begin{figure*}
  \centering
  \subfigure[$\Nrho=2.5$]{
    \includegraphics[width=\mafigME,trim= 150 0 100
    0,clip=true]{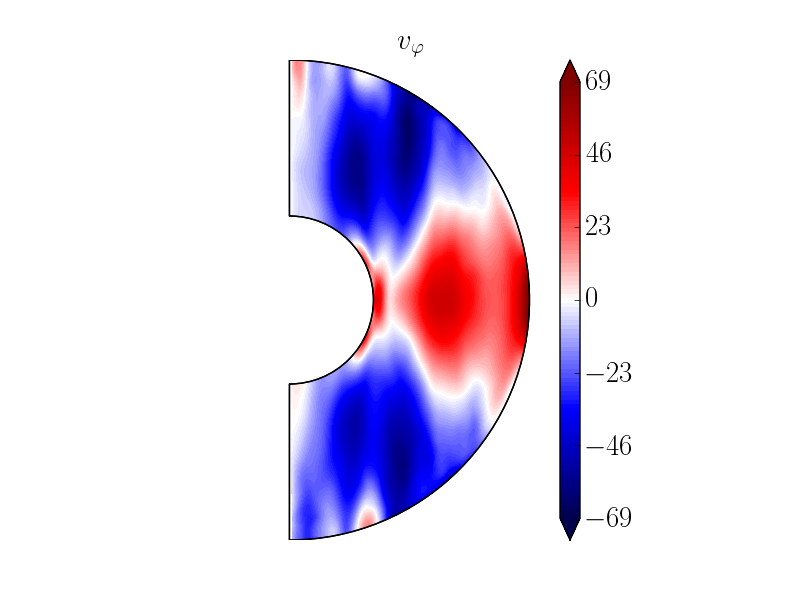}\label{sf:vp25}
  }
  \subfigure[$\Nrho=2.5$]{
    \includegraphics[width=\mafigME,trim= 150 0 100
    0,clip=true]{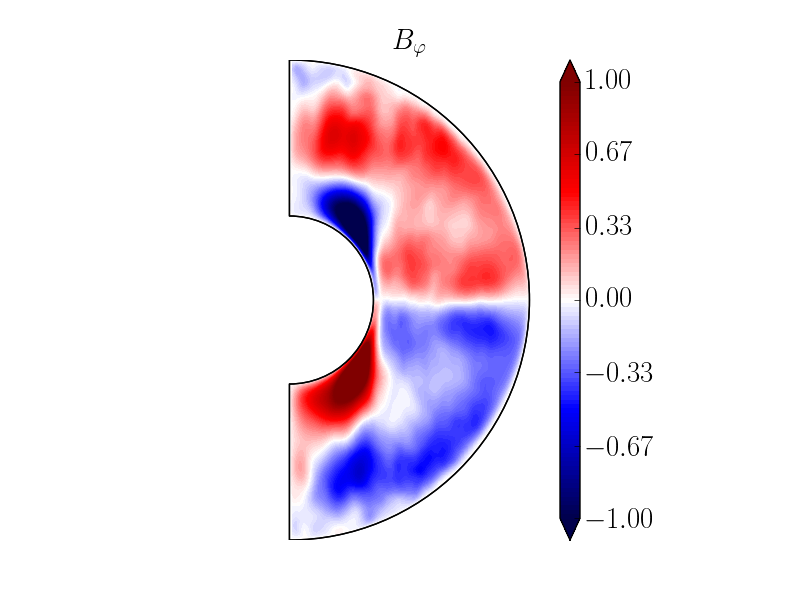}\label{sf:bp25}
  }
  \subfigure[$\Nrho=2.5$]{
    \includegraphics[width=\mafigME,trim= 150 0 100
      0,clip=true]{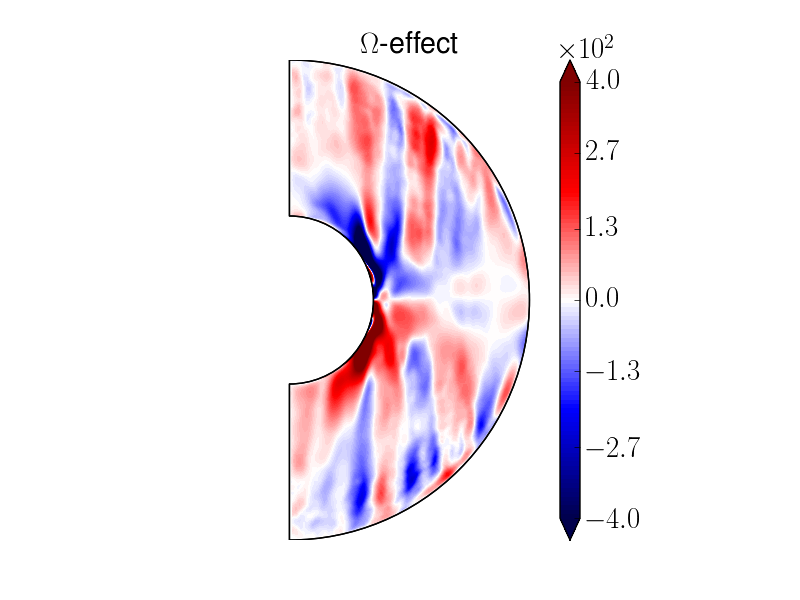}\label{sf:om25}
  }
  \caption{Time-averaged axisymmetric component of the azimuthal
    magnetic field (b) and velocity field (a) for a dipolar
    dynamo with $\Nrho=2.5$, $\Ra=7.40\times10^{6}$,
    $\Pm=2$. }\label{f:npa25}
\end{figure*}
Finally, we try to investigate whether the dynamo mechanisms at work
on the dipolar branch are modified when the stratification is
increased.  We see in Fig.~\ref{f:npa15} that the axisymmetric
azimuthal magnetic field we observe at $\Nrho=1.5$ is strongly
reminiscent of the magnetic structures that can be observed with
Boussinesq models, which are usually interpreted in terms of
$\alpha^2$ dynamos
\citep{olson99,schrinner07,schrinner11a,schrinner12}. Inside the
tangent cylinder, the azimuthal magnetic field is mainly produced by
the $\Omega$-effect, $r \overline{\vect{B}}_r \partial \left(r^{-1}
\overline{\vect{V}}_\varphi\right)/\partial r + r^{-1} \sin \theta
\overline{\vect{B}}_\theta
\partial\left({\sin\theta}^{-1}\overline{\vect{V}}_\varphi
\right)/\partial \theta$, which correlates inside the tangent cylinder
with the axisymmetric azimuthal magnetic field, when comparing
Figs~\ref{sf:bp15} and \ref{sf:om15}. However, outside the tangent
cylinder, the most part of the mean azimuthal field does not seem to
be the result of the $\Omega$-effect, and it is thus likely that the
essential regeneration of the poloidal field is achieved by
$\alpha$-effect, leading to the emergence of characteristic equatorial
patches of opposite polarity \citep[see
  e.g.][]{christensen2011,schrinner12}.  We did not find in our sample
of models tangible evidence that would invalidate this scenario at
higher $\Nrho$. For instance, at $\Nrho=2.5$, we see in
Fig.~\ref{f:npa25} that the major differences lie in the stronger
axisymmetric azimuthal velocity (compare Figs~\ref{sf:vp15} and
\ref{sf:vp25}). Nevertheless, the axisymmetric azimuthal magnetic
field shown in Fig.~\ref{sf:bp25} seems only modified about a
colatitude $\theta \sim \pi/4$ close to the outer surface, and keeps
now the same polarity in each hemisphere outside the tangent cylinder.
This change can be correlated to the modifications of the axisymmetric
azimuthal velocity, which in turn affect the $\Omega$-effect (compare
Figs~\ref{sf:om15} and \ref{sf:om25}). Of course, we are for now
limited to the observation of correlations, but it would be
interesting to have a further insight into the dynamo mechanism in
anelastic simulations using a test field method, in the spirit of the
Boussinesq study by \citet{schrinner12}.

\section{Conclusion}
With this systematic parameter study, we are able to improve our
understanding of the successive modifications that are exhibited by
the stability domain of the dipolar branch when increasing the density
stratification in anelastic dynamo models. In general, dipolar dynamos
are found closer to the onset of convection. Moreover, we show that
dipole-dominated solutions can be observed even at high density
stratifications, provided high enough magnetic Prandtl numbers are
considered. Besides, this study also highlights why dipolar dynamos
seem more difficult to find in anelastic simulations. Indeed, this
tendency is usually reported as a general statement, but here we show
that this impression mainly results from the fact that the dipolar
branch extends on a smaller and smaller range of Rayleigh numbers as
\Nrho{} is increased.  However, despite the relative shrinking of the
stability domain, we found that the critical magnetic Reynolds number
of the dipolar branch seems scarcely modified in the overall
process. At the same time, the higher \Nrho{}, the faster convection
will develop as we depart from the onset. In consequence, the higher
\Nrho{}, the faster is reached the critical Rossby number above which
inertia causes the collapse of the dipole. This explains why dipolar
dynamos become clearly confined in a smaller region of the parameter
space. However, we stress that, in terms of magnetic Reynolds number,
the dynamo threshold does not significantly increase with the density
stratification in the range of \Nrho{} we investigated.

In addition, this study also suggests that the scarcity of dipolar
solutions for substantial density stratifications would thus rather
come from the restriction of the parameter space being currently
explored (because of computational limitations), rather than an
intrinsic modification of the dynamo mechanisms that would be caused
by the density stratification.  Furthermore, if we decrease the Ekman
number from $\Ekman=10^{-4}$ to $3\times 10^{-5}$ keeping
$\Prandtl=1$, we find that we recover three examples of bistable pairs
at $\Nrho=2$, for $\Pm=1$ at $\Ra/\Rac= 2.6$ and for $\Pm \in \lbrace
1,2 \rbrace$ at $\Ra/\Rac= 2.9$. Then, beyond the results of this
study, and for low values of the Ekman number that are currently very
expensive to simulate, it seems more likely that dipolar solutions
will persist in a larger region of parameter space \citep[see
  also][]{duarte,jones2014}.

Despite the fact that it is not straightforward to relate the output
of numerical models with observations \citep{gastineMorin2012}, the
bistability that is reported for numerical simulations can be
similarly observed with real objects. For instance, in a
spectropolarimetric survey done with a sample of active M dwarfs,
\citet{morinD2010} report two distinct categories of magnetic
topologies.  They distinguish strong axisymmetric dipolar fields and
weak fields with significant non-axisymmetric components, and both
configurations seem to be observed on objects with similar stellar
parameters.  After \citet{schrinner12}, we show that the bistable
behaviour observed in numerical models could be a possible way towards
a better understanding of the broad diversity of the magnetic fields
of M dwarfs, and that it cannot be ruled out even when taking into
account the density stratification. The understanding of the impact of
the stratification on the dynamo mechanisms deserves further studies.

\section*{Acknowledgements}
The authors thank L\'{u}cia D. V. Duarte for a thoughtful
review.  This work was granted access to the HPC resources of MesoPSL
financed by the Région \^{I}le-de-France and the project Equip@Meso
(reference ANR-10-EQPX-29-01) of the programme Investissements
d'Avenir supervised by the Agence Nationale pour la
Recherche. Numerical simulations were also carried out at CEMAG and
TGCC computing centres (GENCI project x2013046698). L.~P. acknowledges
financial support from “Programme National de Physique Stellaire”
(PNPS) of CNRS/INSU, France.

\bibliographystyle{mn2eTER}
\bibliography{schrinner,raph}

\begin{thebibliography}{33}
\expandafter\ifx\csname natexlab\endcsname\relax\def\natexlab#1{#1}\fi

\bibitem[{{Alboussi{\`e}re} \& {Ricard}(2013)}]{alboussiere13}
{Alboussi{\`e}re} T., {Ricard} Y., 2013, Journal of Fluid Mechanics, 725, 1

\bibitem[{{Anufriev}, {Jones} \& {Soward}(2005){Anufriev}, {Jones}, \&
  {Soward}}]{anufriev2005}
{Anufriev} A.~P., {Jones} C.~A., {Soward} A.~M., 2005, Physics of the Earth and
  Planetary Interiors, 152, 163

\bibitem[{{Berkoff}, {Kersale} \& {Tobias}(2010){Berkoff}, {Kersale}, \&
  {Tobias}}]{berkoff2010}
{Berkoff} N.~A., {Kersale} E., {Tobias} S.~M., 2010, Geophysical and
  Astrophysical Fluid Dynamics, 104, 545

\bibitem[{{Braginsky} \& {Roberts}(1995)}]{braginsky95}
{Braginsky} S.~I., {Roberts} P.~H., 1995, Geophysical and Astrophysical Fluid
  Dynamics, 79, 1

\bibitem[{{Brown}, {Vasil} \& {Zweibel}(2012){Brown}, {Vasil}, \&
  {Zweibel}}]{brown12}
{Brown} B.~P., {Vasil} G.~M., {Zweibel} E.~G., 2012, \apj, 756, 109

\bibitem[{Christensen(2011)}]{christensen2011}
Christensen U.~R., 2011, Physics of the Earth and Planetary Interiors, 187, 157
  , special Issue: Planetary Magnetism, Dynamo and Dynamics

\bibitem[{{Christensen} \& {Aubert}(2006)}]{christensen06}
{Christensen} U.~R., {Aubert} J., 2006, Geophy. J. Int., 166, 97

\bibitem[{{Donati} \& {Landstreet}(2009)}]{donati09}
{Donati} J.-F., {Landstreet} J.~D., 2009, \araa, 47, 333

\bibitem[{{Dormy}, {Cardin} \& {Jault}(1998){Dormy}, {Cardin}, \&
  {Jault}}]{dormy98}
{Dormy} E., {Cardin} P., {Jault} D., 1998, Earth Planet. Sci. Lett., 160, 15

\bibitem[{{Duarte}(2014)}]{duarte}
{Duarte} L., 2014, PhD thesis, Technische Universität Braunschweig

\bibitem[{{Duarte}, {Gastine} \& {Wicht}(2013){Duarte}, {Gastine}, \&
  {Wicht}}]{duarte2013}
{Duarte} L.~D.~V., {Gastine} T., {Wicht} J., 2013, Physics of the Earth and
  Planetary Interiors, 222, 22

\bibitem[{{Gastine}, {Duarte} \& {Wicht}(2012){Gastine}, {Duarte}, \&
  {Wicht}}]{gastine12}
{Gastine} T., {Duarte} L., {Wicht} J., 2012, \aap, 546, A19

\bibitem[{{Gastine} {et~al}\mbox{.}(2013){Gastine}, {Morin}, {Duarte},
  {Reiners}, {Christensen}, \& {Wicht}}]{gastineMorin2012}
{Gastine} T., {Morin} J., {Duarte} L., {Reiners} A., {Christensen} U.~R.,
  {Wicht} J., 2013, \aap, 549, L5

\bibitem[{{Gastine} \& {Wicht}(2012)}]{gastine12a}
{Gastine} T., {Wicht} J., 2012, Icarus, 219, 428

\bibitem[{Gastine {et~al}\mbox{.}(2014)Gastine, Wicht, Duarte, Heimpel, \&
  Becker}]{gastine2014}
Gastine T., Wicht J., Duarte L. D.~V., Heimpel M., Becker A., 2014, Geophysical
  Research Letters, 41, 5410

\bibitem[{{Gilman} \& {Glatzmaier}(1981)}]{gilman81}
{Gilman} P.~A., {Glatzmaier} G.~A., 1981, \apjs, 45, 335

\bibitem[{{Glatzmaier} \& {Roberts}(1995)}]{glatzroberts95}
{Glatzmaier} G.~A., {Roberts} P.~H., 1995, Nature, 377, 203

\bibitem[{{Gough}(1969)}]{gough69}
{Gough} D.~O., 1969, Journal of Atmospheric Sciences, 26, 448

\bibitem[{Jones(2014)}]{jones2014}
Jones C., 2014, Icarus, 241, 148

\bibitem[{{Jones} {et~al}\mbox{.}(2011){Jones}, {Boronski}, {Brun},
  {Glatzmaier}, {Gastine}, {Miesch}, \& {Wicht}}]{jones11}
{Jones} C.~A., {Boronski} P., {Brun} A.~S., {Glatzmaier} G.~A., {Gastine} T.,
  {Miesch} M.~S., {Wicht} J., 2011, Icarus, 216, 120

\bibitem[{{Jones}, {Kuzanyan} \& {Mitchell}(2009){Jones}, {Kuzanyan}, \&
  {Mitchell}}]{jones_lin}
{Jones} C.~A., {Kuzanyan} K.~M., {Mitchell} R.~H., 2009, Journal of Fluid
  Mechanics, 634, 291

\bibitem[{{Lantz} \& {Fan}(1999)}]{lantz99}
{Lantz} S.~R., {Fan} Y., 1999, \apjs, 121, 247

\bibitem[{Larmor(1919)}]{larmor1919}
Larmor J., 1919, Report of the British Association for the Advancement of
  Science, 87th Meeting, 159

\bibitem[{{Morin} {et~al}\mbox{.}(2010){Morin}, {Donati}, {Petit}, {Delfosse},
  {Forveille}, \& {Jardine}}]{morinD2010}
{Morin} J., {Donati} J.-F., {Petit} P., {Delfosse} X., {Forveille} T.,
  {Jardine} M.~M., 2010, \mnras, 407, 2269

\bibitem[{{Ogura} \& {Phillips}(1962)}]{ogura62}
{Ogura} Y., {Phillips} N.~A., 1962, Journal of Atmospheric Sciences, 19, 173

\bibitem[{{Olson}, {Christensen} \& {Glatzmaier}(1999){Olson}, {Christensen},
  \& {Glatzmaier}}]{olson99}
{Olson} P., {Christensen} U.~R., {Glatzmaier} G.~A., 1999, \jgr, 104, 10383

\bibitem[{{Raynaud}, {Petitdemange} \& {Dormy}(2014){Raynaud}, {Petitdemange},
  \& {Dormy}}]{raynaud2014}
{Raynaud} R., {Petitdemange} L., {Dormy} E., 2014, \aap, 567, A107

\bibitem[{{Sasaki} {et~al}\mbox{.}(2011){Sasaki}, {Takehiro}, {Kuramoto}, \&
  {Hayashi}}]{sasaki2011}
{Sasaki} Y., {Takehiro} S.-i., {Kuramoto} K., {Hayashi} Y.-Y., 2011, Physics of
  the Earth and Planetary Interiors, 188, 203

\bibitem[{{Schrinner}, {Petitdemange} \& {Dormy}(2011){Schrinner},
  {Petitdemange}, \& {Dormy}}]{schrinner11a}
{Schrinner} M., {Petitdemange} L., {Dormy} E., 2011, \aap, 530, A140

\bibitem[{{Schrinner}, {Petitdemange} \& {Dormy}(2012){Schrinner},
  {Petitdemange}, \& {Dormy}}]{schrinner12}
{Schrinner} M., {Petitdemange} L., {Dormy} E., 2012, \apj, 752, 121

\bibitem[{{Schrinner} {et~al}\mbox{.}(2014){Schrinner}, {Petitdemange},
  {Raynaud}, \& {Dormy}}]{schrinner2014}
{Schrinner} M., {Petitdemange} L., {Raynaud} R., {Dormy} E., 2014, \aap, 564,
  A78

\bibitem[{{Schrinner} {et~al}\mbox{.}(2007){Schrinner}, {R{\"a}dler},
  {Schmitt}, {Rheinhardt}, \& {Christensen}}]{schrinner07}
{Schrinner} M., {R{\"a}dler} K.-H., {Schmitt} D., {Rheinhardt} M.,
  {Christensen} U.~R., 2007, Geophys. Astrophys. Fluid Dyn., 101, 81

\bibitem[{Yadav {et~al}\mbox{.}(2013)Yadav, Gastine, Christensen, \&
  Duarte}]{yadav13}
Yadav R.~K., Gastine T., Christensen U.~R., Duarte L. D.~V., 2013, The
  Astrophysical Journal, 774, 6

\end{thebibliography}

\appendix

\section{Numerical models}\label{s:appendix}
\onecolumn
\begin{longtable}{ccccccccc}
\caption{Overview of the simulations carried out, with
  $\Ekman=10^{-4}$, $\Prandtl=1$, $\aspectratio=0.35$, and $n=2$.}\label{table}\\
\hline\hline
model &$ \Nrho $&$ \Ra $&$ \Pm $&$ \Ro $&$ \Rol $&$ \Roz $&$ \Lo $&$ \fdipAX $\\
\hline
\endfirsthead
\caption{(continued)}\\
\hline\hline
model &$ \Nrho $&$ \Ra $&$ \Pm $&$ \Ro $&$ \Rol $&$ \Roz $&$ \Lo $&$ \fdipAX $\\
\hline
\endhead
\hline
\endfoot
$ \texttt{001}\mm $ & $ 0.5 $ & $ 1.500 \times 10^{6} $ & $ 2.00 $ & $ 3.2 \times 10^{-3} $ & $ 2.0 \times 10^{-2} $ & $ 1.8 \times 10^{-3} $ & $ 2.3 \times 10^{-3} $ & $ 6.6 \times 10^{-2} $ \\
$ \texttt{002}\mm $ & $ 0.5 $ & $ 1.750 \times 10^{6} $ & $ 1.00 $ & $ 3.7 \times 10^{-3} $ & $ 2.2 \times 10^{-2} $ & $ 2.7 \times 10^{-3} $ & $ 1.8 \times 10^{-3} $ & $ 1.6 \times 10^{-2} $ \\
$ \texttt{003}\mm $ & $ 0.5 $ & $ 1.800 \times 10^{6} $ & $ 0.75 $ & $ 3.5 \times 10^{-3} $ & $ 2.8 \times 10^{-2} $ & $ 2.8 \times 10^{-3} $ & $ 1.9 \times 10^{-3} $ & $ 2.2 \times 10^{-3} $ \\
$ \texttt{004}\mm $ & $ 0.5 $ & $ 1.850 \times 10^{6} $ & $ 1.00 $ & $ 3.9 \times 10^{-3} $ & $ 2.4 \times 10^{-2} $ & $ 4.2 \times 10^{-3} $ & $ 2.0 \times 10^{-3} $ & $ 2.5 \times 10^{-2} $ \\
$ \texttt{005}\mm $ & $ 0.5 $ & $ 2.000 \times 10^{6} $ & $ 0.75 $ & $ 4.1 \times 10^{-3} $ & $ 2.3 \times 10^{-2} $ & $ 5.2 \times 10^{-3} $ & $ 2.4 \times 10^{-3} $ & $ 9.0 \times 10^{-5} $ \\
$ \texttt{006}\mm $ & $ 0.5 $ & $ 2.000 \times 10^{6} $ & $ 1.50 $ & $ 5.5 \times 10^{-3} $ & $ 2.8 \times 10^{-2} $ & $ 3.0 \times 10^{-3} $ & $ 3.1 \times 10^{-3} $ & $ 5.3 \times 10^{-2} $ \\
$ \texttt{007}\mm $ & $ 0.5 $ & $ 2.000 \times 10^{6} $ & $ 2.00 $ & $ 5.3 \times 10^{-3} $ & $ 2.7 \times 10^{-2} $ & $ 2.0 \times 10^{-3} $ & $ 3.7 \times 10^{-3} $ & $ 9.4 \times 10^{-2} $ \\
$ \texttt{008}\mm $ & $ 0.5 $ & $ 2.000 \times 10^{6} $ & $ 3.00 $ & $ 5.1 \times 10^{-3} $ & $ 2.7 \times 10^{-2} $ & $ 1.7 \times 10^{-3} $ & $ 4.1 \times 10^{-3} $ & $ 2.7 \times 10^{-1} $ \\
$ \texttt{008}\dd $ & $ 0.5 $ & $ 2.000 \times 10^{6} $ & $ 3.00 $ & $ 4.8 \times 10^{-3} $ & $ 2.6 \times 10^{-2} $ & $ 1.0 \times 10^{-3} $ & $ 5.5 \times 10^{-3} $ & $ 7.7 \times 10^{-1} $ \\
$ \texttt{009}\dd $ & $ 0.5 $ & $ 2.000 \times 10^{6} $ & $ 5.00 $ & $ 4.8 \times 10^{-3} $ & $ 2.5 \times 10^{-2} $ & $ 1.1 \times 10^{-3} $ & $ 5.7 \times 10^{-3} $ & $ 6.8 \times 10^{-1} $ \\
$ \texttt{010}\mm $ & $ 0.5 $ & $ 2.500 \times 10^{6} $ & $ 0.75 $ & $ 7.7 \times 10^{-3} $ & $ 3.3 \times 10^{-2} $ & $ 5.2 \times 10^{-3} $ & $ 3.8 \times 10^{-3} $ & $ 5.1 \times 10^{-2} $ \\
$ \texttt{011}\mm $ & $ 0.5 $ & $ 2.500 \times 10^{6} $ & $ 1.00 $ & $ 7.2 \times 10^{-3} $ & $ 3.5 \times 10^{-2} $ & $ 3.6 \times 10^{-3} $ & $ 4.5 \times 10^{-3} $ & $ 7.6 \times 10^{-2} $ \\
$ \texttt{012}\mm $ & $ 0.5 $ & $ 2.500 \times 10^{6} $ & $ 1.50 $ & $ 7.1 \times 10^{-3} $ & $ 3.5 \times 10^{-2} $ & $ 3.1 \times 10^{-3} $ & $ 4.5 \times 10^{-3} $ & $ 3.3 \times 10^{-1} $ \\
$ \texttt{013}\dd $ & $ 0.5 $ & $ 2.500 \times 10^{6} $ & $ 4.00 $ & $ 6.0 \times 10^{-3} $ & $ 3.3 \times 10^{-2} $ & $ 1.5 \times 10^{-3} $ & $ 6.9 \times 10^{-3} $ & $ 7.3 \times 10^{-1} $ \\
$ \texttt{014}\mm $ & $ 0.5 $ & $ 3.000 \times 10^{6} $ & $ 1.00 $ & $ 8.6 \times 10^{-3} $ & $ 4.2 \times 10^{-2} $ & $ 4.0 \times 10^{-3} $ & $ 5.5 \times 10^{-3} $ & $ 1.3 \times 10^{-1} $ \\
$ \texttt{015}\mm $ & $ 0.5 $ & $ 3.000 \times 10^{6} $ & $ 2.00 $ & $ 8.2 \times 10^{-3} $ & $ 4.2 \times 10^{-2} $ & $ 2.8 \times 10^{-3} $ & $ 6.3 \times 10^{-3} $ & $ 3.0 \times 10^{-1} $ \\
$ \texttt{015}\dd $ & $ 0.5 $ & $ 3.000 \times 10^{6} $ & $ 2.00 $ & $ 7.5 \times 10^{-3} $ & $ 3.9 \times 10^{-2} $ & $ 1.7 \times 10^{-3} $ & $ 7.9 \times 10^{-3} $ & $ 7.5 \times 10^{-1} $ \\
$ \texttt{016}\dd $ & $ 0.5 $ & $ 3.000 \times 10^{6} $ & $ 3.00 $ & $ 7.5 \times 10^{-3} $ & $ 4.0 \times 10^{-2} $ & $ 1.7 \times 10^{-3} $ & $ 8.5 \times 10^{-3} $ & $ 7.2 \times 10^{-1} $ \\
$ \texttt{017}\dd $ & $ 0.5 $ & $ 3.000 \times 10^{6} $ & $ 4.00 $ & $ 7.5 \times 10^{-3} $ & $ 4.0 \times 10^{-2} $ & $ 1.7 \times 10^{-3} $ & $ 8.5 \times 10^{-3} $ & $ 7.2 \times 10^{-1} $ \\
$ \texttt{018}\dd $ & $ 0.5 $ & $ 3.000 \times 10^{6} $ & $ 5.00 $ & $ 7.5 \times 10^{-3} $ & $ 4.1 \times 10^{-2} $ & $ 1.8 \times 10^{-3} $ & $ 8.8 \times 10^{-3} $ & $ 6.8 \times 10^{-1} $ \\
$ \texttt{019}\dd $ & $ 0.5 $ & $ 3.000 \times 10^{6} $ & $ 6.00 $ & $ 7.6 \times 10^{-3} $ & $ 4.1 \times 10^{-2} $ & $ 1.9 \times 10^{-3} $ & $ 9.0 \times 10^{-3} $ & $ 6.4 \times 10^{-1} $ \\
$ \texttt{020}\mm $ & $ 0.5 $ & $ 4.000 \times 10^{6} $ & $ 0.50 $ & $ 1.3 \times 10^{-2} $ & $ 5.3 \times 10^{-2} $ & $ 9.3 \times 10^{-3} $ & $ 6.6 \times 10^{-3} $ & $ 2.3 \times 10^{-1} $ \\
$ \texttt{021}\mm $ & $ 0.5 $ & $ 4.000 \times 10^{6} $ & $ 1.00 $ & $ 1.2 \times 10^{-2} $ & $ 5.6 \times 10^{-2} $ & $ 5.7 \times 10^{-3} $ & $ 7.9 \times 10^{-3} $ & $ 2.8 \times 10^{-1} $ \\
$ \texttt{021}\dd $ & $ 0.5 $ & $ 4.000 \times 10^{6} $ & $ 1.00 $ & $ 1.0 \times 10^{-2} $ & $ 5.4 \times 10^{-2} $ & $ 2.0 \times 10^{-3} $ & $ 1.1 \times 10^{-2} $ & $ 8.7 \times 10^{-1} $ \\
$ \texttt{022}\mm $ & $ 0.5 $ & $ 4.000 \times 10^{6} $ & $ 2.00 $ & $ 1.5 \times 10^{-2} $ & $ 5.7 \times 10^{-2} $ & $ 3.5 \times 10^{-3} $ & $ 8.8 \times 10^{-3} $ & $ 2.6 \times 10^{-1} $ \\
$ \texttt{022}\dd $ & $ 0.5 $ & $ 4.000 \times 10^{6} $ & $ 2.00 $ & $ 1.0 \times 10^{-2} $ & $ 5.4 \times 10^{-2} $ & $ 1.9 \times 10^{-3} $ & $ 1.1 \times 10^{-2} $ & $ 6.6 \times 10^{-1} $ \\
$ \texttt{023}\dd $ & $ 0.5 $ & $ 4.000 \times 10^{6} $ & $ 3.00 $ & $ 1.0 \times 10^{-2} $ & $ 5.3 \times 10^{-2} $ & $ 2.0 \times 10^{-3} $ & $ 1.1 \times 10^{-2} $ & $ 6.8 \times 10^{-1} $ \\
$ \texttt{024}\dd $ & $ 0.5 $ & $ 4.000 \times 10^{6} $ & $ 4.00 $ & $ 1.0 \times 10^{-2} $ & $ 5.3 \times 10^{-2} $ & $ 1.9 \times 10^{-3} $ & $ 1.1 \times 10^{-2} $ & $ 6.5 \times 10^{-1} $ \\
$ \texttt{025}\dd $ & $ 0.5 $ & $ 4.000 \times 10^{6} $ & $ 6.00 $ & $ 1.0 \times 10^{-2} $ & $ 5.3 \times 10^{-2} $ & $ 2.2 \times 10^{-3} $ & $ 1.2 \times 10^{-2} $ & $ 5.4 \times 10^{-1} $ \\
$ \texttt{026}\mm $ & $ 0.5 $ & $ 5.000 \times 10^{6} $ & $ 0.50 $ & $ 1.5 \times 10^{-2} $ & $ 6.8 \times 10^{-2} $ & $ 8.4 \times 10^{-3} $ & $ 8.8 \times 10^{-3} $ & $ 2.3 \times 10^{-1} $ \\
$ \texttt{027}\mm $ & $ 0.5 $ & $ 5.000 \times 10^{6} $ & $ 1.00 $ & $ 1.4 \times 10^{-2} $ & $ 7.1 \times 10^{-2} $ & $ 5.5 \times 10^{-3} $ & $ 1.0 \times 10^{-2} $ & $ 2.5 \times 10^{-1} $ \\
$ \texttt{027}\dd $ & $ 0.5 $ & $ 5.000 \times 10^{6} $ & $ 1.00 $ & $ 1.3 \times 10^{-2} $ & $ 6.7 \times 10^{-2} $ & $ 2.0 \times 10^{-3} $ & $ 1.4 \times 10^{-2} $ & $ 8.1 \times 10^{-1} $ \\
$ \texttt{028}\dd $ & $ 0.5 $ & $ 5.000 \times 10^{6} $ & $ 2.00 $ & $ 1.3 \times 10^{-2} $ & $ 6.6 \times 10^{-2} $ & $ 2.2 \times 10^{-3} $ & $ 1.4 \times 10^{-2} $ & $ 6.8 \times 10^{-1} $ \\
$ \texttt{029}\dd $ & $ 0.5 $ & $ 5.000 \times 10^{6} $ & $ 3.00 $ & $ 1.3 \times 10^{-2} $ & $ 6.8 \times 10^{-2} $ & $ 2.1 \times 10^{-3} $ & $ 1.4 \times 10^{-2} $ & $ 7.2 \times 10^{-1} $ \\
$ \texttt{030}\dd $ & $ 0.5 $ & $ 5.000 \times 10^{6} $ & $ 4.00 $ & $ 1.3 \times 10^{-2} $ & $ 6.5 \times 10^{-2} $ & $ 2.5 \times 10^{-3} $ & $ 1.6 \times 10^{-2} $ & $ 7.0 \times 10^{-1} $ \\
$ \texttt{031}\dd $ & $ 0.5 $ & $ 5.000 \times 10^{6} $ & $ 5.00 $ & $ 1.3 \times 10^{-2} $ & $ 6.7 \times 10^{-2} $ & $ 2.5 \times 10^{-3} $ & $ 1.5 \times 10^{-2} $ & $ 6.8 \times 10^{-1} $ \\
$ \texttt{032}\mm $ & $ 0.5 $ & $ 6.000 \times 10^{6} $ & $ 1.00 $ & $ 1.7 \times 10^{-2} $ & $ 8.5 \times 10^{-2} $ & $ 6.3 \times 10^{-3} $ & $ 1.3 \times 10^{-2} $ & $ 2.8 \times 10^{-1} $ \\
$ \texttt{032}\dd $ & $ 0.5 $ & $ 6.000 \times 10^{6} $ & $ 1.00 $ & $ 1.5 \times 10^{-2} $ & $ 8.0 \times 10^{-2} $ & $ 2.6 \times 10^{-3} $ & $ 1.7 \times 10^{-2} $ & $ 8.2 \times 10^{-1} $ \\
$ \texttt{033}\mm $ & $ 0.5 $ & $ 7.000 \times 10^{6} $ & $ 1.00 $ & $ 2.0 \times 10^{-2} $ & $ 9.6 \times 10^{-2} $ & $ 7.4 \times 10^{-3} $ & $ 1.4 \times 10^{-2} $ & $ 2.5 \times 10^{-1} $ \\
$ \texttt{033}\dd $ & $ 0.5 $ & $ 7.000 \times 10^{6} $ & $ 1.00 $ & $ 1.8 \times 10^{-2} $ & $ 8.6 \times 10^{-2} $ & $ 3.0 \times 10^{-3} $ & $ 2.0 \times 10^{-2} $ & $ 8.5 \times 10^{-1} $ \\
$ \texttt{034}\mm $ & $ 0.5 $ & $ 9.000 \times 10^{6} $ & $ 1.00 $ & $ 2.5 \times 10^{-2} $ & $ 1.2 \times 10^{-1} $ & $ 8.2 \times 10^{-3} $ & $ 1.9 \times 10^{-2} $ & $ 3.6 \times 10^{-1} $ \\
$ \texttt{035}\mm $ & $ 0.5 $ & $ 1.000 \times 10^{7} $ & $ 1.00 $ & $ 2.7 \times 10^{-2} $ & $ 1.3 \times 10^{-1} $ & $ 8.7 \times 10^{-3} $ & $ 2.1 \times 10^{-2} $ & $ 3.4 \times 10^{-1} $ \\
$ \texttt{036}\mm $ & $ 1.5 $ & $ 2.500 \times 10^{6} $ & $ 0.75 $ & $ 3.9 \times 10^{-3} $ & $ 2.6 \times 10^{-2} $ & $ 2.9 \times 10^{-3} $ & $ 3.3 \times 10^{-3} $ & $ 4.7 \times 10^{-2} $ \\
$ \texttt{037}\mm $ & $ 1.5 $ & $ 2.500 \times 10^{6} $ & $ 1.00 $ & $ 4.5 \times 10^{-3} $ & $ 2.7 \times 10^{-2} $ & $ 2.0 \times 10^{-3} $ & $ 3.6 \times 10^{-3} $ & $ 1.9 \times 10^{-1} $ \\
$ \texttt{038}\mm $ & $ 1.5 $ & $ 2.500 \times 10^{6} $ & $ 1.50 $ & $ 4.3 \times 10^{-3} $ & $ 2.8 \times 10^{-2} $ & $ 9.2 \times 10^{-4} $ & $ 4.4 \times 10^{-3} $ & $ 1.0 \times 10^{-1} $ \\
$ \texttt{038}\dd $ & $ 1.5 $ & $ 2.500 \times 10^{6} $ & $ 1.50 $ & $ 4.3 \times 10^{-3} $ & $ 2.8 \times 10^{-2} $ & $ 8.2 \times 10^{-4} $ & $ 5.0 \times 10^{-3} $ & $ 7.9 \times 10^{-1} $ \\
$ \texttt{039}\dd $ & $ 1.5 $ & $ 2.500 \times 10^{6} $ & $ 2.00 $ & $ 4.3 \times 10^{-3} $ & $ 3.0 \times 10^{-2} $ & $ 7.7 \times 10^{-4} $ & $ 4.6 \times 10^{-3} $ & $ 7.1 \times 10^{-1} $ \\
$ \texttt{040}\dd $ & $ 1.5 $ & $ 2.500 \times 10^{6} $ & $ 3.00 $ & $ 4.2 \times 10^{-3} $ & $ 2.9 \times 10^{-2} $ & $ 8.0 \times 10^{-4} $ & $ 4.8 \times 10^{-3} $ & $ 7.3 \times 10^{-1} $ \\
$ \texttt{041}\mm $ & $ 1.5 $ & $ 3.000 \times 10^{6} $ & $ 0.75 $ & $ 5.5 \times 10^{-3} $ & $ 3.6 \times 10^{-2} $ & $ 2.9 \times 10^{-3} $ & $ 4.8 \times 10^{-3} $ & $ 2.1 \times 10^{-1} $ \\
$ \texttt{042}\mm $ & $ 1.5 $ & $ 3.000 \times 10^{6} $ & $ 1.00 $ & $ 6.0 \times 10^{-3} $ & $ 3.6 \times 10^{-2} $ & $ 2.0 \times 10^{-3} $ & $ 5.1 \times 10^{-3} $ & $ 3.9 \times 10^{-1} $ \\
$ \texttt{042}\dd $ & $ 1.5 $ & $ 3.000 \times 10^{6} $ & $ 1.00 $ & $ 5.7 \times 10^{-3} $ & $ 3.7 \times 10^{-2} $ & $ 1.1 \times 10^{-3} $ & $ 6.4 \times 10^{-3} $ & $ 8.5 \times 10^{-1} $ \\
$ \texttt{043}\dd $ & $ 1.5 $ & $ 3.000 \times 10^{6} $ & $ 2.00 $ & $ 5.5 \times 10^{-3} $ & $ 3.7 \times 10^{-2} $ & $ 1.1 \times 10^{-3} $ & $ 6.2 \times 10^{-3} $ & $ 7.1 \times 10^{-1} $ \\
$ \texttt{044}\dd $ & $ 1.5 $ & $ 3.700 \times 10^{6} $ & $ 3.00 $ & $ 7.3 \times 10^{-3} $ & $ 4.7 \times 10^{-2} $ & $ 1.4 \times 10^{-3} $ & $ 7.9 \times 10^{-3} $ & $ 6.0 \times 10^{-1} $ \\
$ \texttt{045}\mm $ & $ 1.5 $ & $ 4.000 \times 10^{6} $ & $ 0.50 $ & $ 9.5 \times 10^{-3} $ & $ 4.9 \times 10^{-2} $ & $ 5.3 \times 10^{-3} $ & $ 6.2 \times 10^{-3} $ & $ 2.8 \times 10^{-1} $ \\
$ \texttt{046}\mm $ & $ 1.5 $ & $ 4.000 \times 10^{6} $ & $ 0.75 $ & $ 8.7 \times 10^{-3} $ & $ 5.2 \times 10^{-2} $ & $ 3.0 \times 10^{-3} $ & $ 6.9 \times 10^{-3} $ & $ 4.5 \times 10^{-1} $ \\
$ \texttt{046}\dd $ & $ 1.5 $ & $ 4.000 \times 10^{6} $ & $ 0.75 $ & $ 8.5 \times 10^{-3} $ & $ 5.3 \times 10^{-2} $ & $ 1.5 \times 10^{-3} $ & $ 9.3 \times 10^{-3} $ & $ 8.2 \times 10^{-1} $ \\
$ \texttt{047}\dd $ & $ 1.5 $ & $ 4.000 \times 10^{6} $ & $ 1.00 $ & $ 8.5 \times 10^{-3} $ & $ 5.2 \times 10^{-2} $ & $ 1.4 \times 10^{-3} $ & $ 9.4 \times 10^{-3} $ & $ 7.8 \times 10^{-1} $ \\
$ \texttt{048}\dd $ & $ 1.5 $ & $ 4.000 \times 10^{6} $ & $ 2.00 $ & $ 8.5 \times 10^{-3} $ & $ 5.3 \times 10^{-2} $ & $ 1.5 \times 10^{-3} $ & $ 8.4 \times 10^{-3} $ & $ 6.9 \times 10^{-1} $ \\
$ \texttt{049}\mm $ & $ 1.5 $ & $ 4.625 \times 10^{6} $ & $ 0.50 $ & $ 9.5 \times 10^{-3} $ & $ 5.8 \times 10^{-2} $ & $ 6.0 \times 10^{-3} $ & $ 7.5 \times 10^{-3} $ & $ 4.5 \times 10^{-1} $ \\
$ \texttt{050}\mm $ & $ 1.5 $ & $ 4.625 \times 10^{6} $ & $ 0.75 $ & $ 1.0 \times 10^{-2} $ & $ 6.2 \times 10^{-2} $ & $ 3.6 \times 10^{-3} $ & $ 8.1 \times 10^{-3} $ & $ 5.3 \times 10^{-1} $ \\
$ \texttt{050}\dd $ & $ 1.5 $ & $ 4.625 \times 10^{6} $ & $ 0.75 $ & $ 1.0 \times 10^{-2} $ & $ 6.2 \times 10^{-2} $ & $ 1.7 \times 10^{-3} $ & $ 1.1 \times 10^{-2} $ & $ 8.6 \times 10^{-1} $ \\
$ \texttt{051}\dd $ & $ 1.5 $ & $ 4.625 \times 10^{6} $ & $ 1.00 $ & $ 1.0 \times 10^{-2} $ & $ 6.3 \times 10^{-2} $ & $ 1.7 \times 10^{-3} $ & $ 1.1 \times 10^{-2} $ & $ 7.6 \times 10^{-1} $ \\
$ \texttt{052}\dd $ & $ 1.5 $ & $ 5.000 \times 10^{6} $ & $ 1.00 $ & $ 1.1 \times 10^{-2} $ & $ 6.7 \times 10^{-2} $ & $ 1.8 \times 10^{-3} $ & $ 1.2 \times 10^{-2} $ & $ 7.5 \times 10^{-1} $ \\
$ \texttt{053}\dd $ & $ 1.5 $ & $ 5.000 \times 10^{6} $ & $ 2.00 $ & $ 1.1 \times 10^{-2} $ & $ 6.7 \times 10^{-2} $ & $ 1.8 \times 10^{-3} $ & $ 1.2 \times 10^{-2} $ & $ 6.5 \times 10^{-1} $ \\
$ \texttt{054}\mm $ & $ 1.5 $ & $ 5.550 \times 10^{6} $ & $ 0.75 $ & $ 1.2 \times 10^{-2} $ & $ 7.4 \times 10^{-2} $ & $ 4.0 \times 10^{-3} $ & $ 9.9 \times 10^{-3} $ & $ 4.7 \times 10^{-1} $ \\
$ \texttt{054}\dd $ & $ 1.5 $ & $ 5.550 \times 10^{6} $ & $ 0.75 $ & $ 1.3 \times 10^{-2} $ & $ 7.7 \times 10^{-2} $ & $ 2.0 \times 10^{-3} $ & $ 1.4 \times 10^{-2} $ & $ 8.4 \times 10^{-1} $ \\
$ \texttt{055}\dd $ & $ 1.5 $ & $ 5.550 \times 10^{6} $ & $ 1.00 $ & $ 1.3 \times 10^{-2} $ & $ 7.7 \times 10^{-2} $ & $ 2.1 \times 10^{-3} $ & $ 1.3 \times 10^{-2} $ & $ 7.4 \times 10^{-1} $ \\
$ \texttt{056}\dd $ & $ 1.5 $ & $ 5.550 \times 10^{6} $ & $ 2.00 $ & $ 1.2 \times 10^{-2} $ & $ 7.4 \times 10^{-2} $ & $ - $ & $ 1.5 \times 10^{-2} $ & $ 6.3 \times 10^{-1} $ \\
$ \texttt{057}\mm $ & $ 1.5 $ & $ 6.500 \times 10^{6} $ & $ 0.50 $ & $ 1.6 \times 10^{-2} $ & $ 8.8 \times 10^{-2} $ & $ 5.5 \times 10^{-3} $ & $ 1.1 \times 10^{-2} $ & $ 4.1 \times 10^{-1} $ \\
$ \texttt{058}\mm $ & $ 1.5 $ & $ 6.500 \times 10^{6} $ & $ 0.75 $ & $ 1.5 \times 10^{-2} $ & $ 8.8 \times 10^{-2} $ & $ 4.3 \times 10^{-3} $ & $ 1.2 \times 10^{-2} $ & $ 4.8 \times 10^{-1} $ \\
$ \texttt{058}\dd $ & $ 1.5 $ & $ 6.500 \times 10^{6} $ & $ 0.75 $ & $ 1.5 \times 10^{-2} $ & $ 8.9 \times 10^{-2} $ & $ 2.4 \times 10^{-3} $ & $ 1.6 \times 10^{-2} $ & $ 8.3 \times 10^{-1} $ \\
$ \texttt{059}\dd $ & $ 1.5 $ & $ 6.500 \times 10^{6} $ & $ 1.00 $ & $ 1.5 \times 10^{-2} $ & $ 8.8 \times 10^{-2} $ & $ 2.5 \times 10^{-3} $ & $ 1.6 \times 10^{-2} $ & $ 7.7 \times 10^{-1} $ \\
$ \texttt{060}\mm $ & $ 1.5 $ & $ 8.000 \times 10^{6} $ & $ 0.75 $ & $ 1.9 \times 10^{-2} $ & $ 1.1 \times 10^{-1} $ & $ 5.4 \times 10^{-3} $ & $ 1.5 \times 10^{-2} $ & $ 3.8 \times 10^{-1} $ \\
$ \texttt{060}\dd $ & $ 1.5 $ & $ 8.000 \times 10^{6} $ & $ 0.75 $ & $ 1.9 \times 10^{-2} $ & $ 1.1 \times 10^{-1} $ & $ 3.3 \times 10^{-3} $ & $ 2.0 \times 10^{-2} $ & $ 8.2 \times 10^{-1} $ \\
$ \texttt{061}\dd $ & $ 1.5 $ & $ 8.000 \times 10^{6} $ & $ 1.00 $ & $ 1.9 \times 10^{-2} $ & $ 1.1 \times 10^{-1} $ & $ 3.0 \times 10^{-3} $ & $ 2.0 \times 10^{-2} $ & $ 7.7 \times 10^{-1} $ \\
$ \texttt{062}\mm $ & $ 1.5 $ & $ 9.000 \times 10^{6} $ & $ 0.50 $ & $ 2.2 \times 10^{-2} $ & $ 1.2 \times 10^{-1} $ & $ 6.4 \times 10^{-3} $ & $ 1.6 \times 10^{-2} $ & $ 4.5 \times 10^{-1} $ \\
$ \texttt{063}\mm $ & $ 1.5 $ & $ 9.000 \times 10^{6} $ & $ 1.00 $ & $ 2.2 \times 10^{-2} $ & $ 1.2 \times 10^{-1} $ & $ 5.1 \times 10^{-3} $ & $ 1.8 \times 10^{-2} $ & $ 3.1 \times 10^{-1} $ \\
$ \texttt{064}\mm $ & $ 1.5 $ & $ 1.000 \times 10^{7} $ & $ 0.50 $ & $ 2.5 \times 10^{-2} $ & $ 1.3 \times 10^{-1} $ & $ 8.0 \times 10^{-3} $ & $ 1.8 \times 10^{-2} $ & $ 3.0 \times 10^{-1} $ \\
$ \texttt{065}\mm $ & $ 2.0 $ & $ 3.000 \times 10^{6} $ & $ 1.00 $ & $ 4.0 \times 10^{-3} $ & $ 2.9 \times 10^{-2} $ & $ 1.8 \times 10^{-3} $ & $ 3.8 \times 10^{-3} $ & $ 2.2 \times 10^{-1} $ \\
$ \texttt{066}\dd $ & $ 2.0 $ & $ 3.000 \times 10^{6} $ & $ 2.00 $ & $ 4.0 \times 10^{-3} $ & $ 3.1 \times 10^{-2} $ & $ 5.5 \times 10^{-4} $ & $ 4.5 \times 10^{-3} $ & $ 7.7 \times 10^{-1} $ \\
$ \texttt{067}\mm $ & $ 2.0 $ & $ 4.000 \times 10^{6} $ & $ 1.00 $ & $ 6.8 \times 10^{-3} $ & $ 4.6 \times 10^{-2} $ & $ 1.7 \times 10^{-3} $ & $ 6.0 \times 10^{-3} $ & $ 2.9 \times 10^{-1} $ \\
$ \texttt{068}\dd $ & $ 2.0 $ & $ 4.000 \times 10^{6} $ & $ 2.00 $ & $ 6.5 \times 10^{-3} $ & $ 4.6 \times 10^{-2} $ & $ 1.1 \times 10^{-3} $ & $ 7.2 \times 10^{-3} $ & $ 7.2 \times 10^{-1} $ \\
$ \texttt{069}\dd $ & $ 2.0 $ & $ 4.000 \times 10^{6} $ & $ 3.00 $ & $ 6.6 \times 10^{-3} $ & $ 4.6 \times 10^{-2} $ & $ 1.1 \times 10^{-3} $ & $ 7.3 \times 10^{-3} $ & $ 6.4 \times 10^{-1} $ \\
$ \texttt{070}\mm $ & $ 2.0 $ & $ 5.000 \times 10^{6} $ & $ 0.50 $ & $ 8.3 \times 10^{-3} $ & $ 5.5 \times 10^{-2} $ & $ 5.4 \times 10^{-3} $ & $ 6.4 \times 10^{-3} $ & $ 2.3 \times 10^{-1} $ \\
$ \texttt{071}\mm $ & $ 2.0 $ & $ 5.000 \times 10^{6} $ & $ 1.00 $ & $ 9.2 \times 10^{-3} $ & $ 6.0 \times 10^{-2} $ & $ 2.1 \times 10^{-3} $ & $ 7.5 \times 10^{-3} $ & $ 2.9 \times 10^{-1} $ \\
$ \texttt{072}\dd $ & $ 2.0 $ & $ 5.000 \times 10^{6} $ & $ 1.50 $ & $ 9.3 \times 10^{-3} $ & $ 6.4 \times 10^{-2} $ & $ 1.6 \times 10^{-3} $ & $ 9.9 \times 10^{-3} $ & $ 6.8 \times 10^{-1} $ \\
$ \texttt{073}\dd $ & $ 2.0 $ & $ 5.000 \times 10^{6} $ & $ 2.00 $ & $ 9.1 \times 10^{-3} $ & $ 6.4 \times 10^{-2} $ & $ 1.7 \times 10^{-3} $ & $ 1.0 \times 10^{-2} $ & $ 6.3 \times 10^{-1} $ \\
$ \texttt{074}\dd $ & $ 2.0 $ & $ 5.000 \times 10^{6} $ & $ 3.00 $ & $ 9.0 \times 10^{-3} $ & $ 6.1 \times 10^{-2} $ & $ 1.5 \times 10^{-3} $ & $ 1.0 \times 10^{-2} $ & $ 6.1 \times 10^{-1} $ \\
$ \texttt{075}\mm $ & $ 2.0 $ & $ 6.000 \times 10^{6} $ & $ 0.50 $ & $ 1.1 \times 10^{-2} $ & $ 7.3 \times 10^{-2} $ & $ 4.1 \times 10^{-3} $ & $ 8.2 \times 10^{-3} $ & $ 2.5 \times 10^{-1} $ \\
$ \texttt{076}\dd $ & $ 2.0 $ & $ 6.000 \times 10^{6} $ & $ 2.00 $ & $ 1.2 \times 10^{-2} $ & $ 7.9 \times 10^{-2} $ & $ 2.0 \times 10^{-3} $ & $ 1.3 \times 10^{-2} $ & $ 5.6 \times 10^{-1} $ \\
$ \texttt{077}\mm $ & $ 2.0 $ & $ 7.000 \times 10^{6} $ & $ 0.70 $ & $ 1.5 \times 10^{-2} $ & $ 8.7 \times 10^{-2} $ & $ 4.0 \times 10^{-3} $ & $ 1.2 \times 10^{-2} $ & $ 1.7 \times 10^{-1} $ \\
$ \texttt{078}\mm $ & $ 2.0 $ & $ 7.000 \times 10^{6} $ & $ 1.00 $ & $ 1.4 \times 10^{-2} $ & $ 9.0 \times 10^{-2} $ & $ 3.0 \times 10^{-3} $ & $ 1.2 \times 10^{-2} $ & $ 4.4 \times 10^{-1} $ \\
$ \texttt{079}\dd $ & $ 2.0 $ & $ 7.000 \times 10^{6} $ & $ 1.50 $ & $ 1.4 \times 10^{-2} $ & $ 9.0 \times 10^{-2} $ & $ 1.9 \times 10^{-3} $ & $ 1.6 \times 10^{-2} $ & $ 5.9 \times 10^{-1} $ \\
$ \texttt{080}\dd $ & $ 2.0 $ & $ 7.000 \times 10^{6} $ & $ 2.00 $ & $ 1.4 \times 10^{-2} $ & $ 8.8 \times 10^{-2} $ & $ 1.7 \times 10^{-3} $ & $ 1.6 \times 10^{-2} $ & $ 6.7 \times 10^{-1} $ \\
$ \texttt{081}\dd $ & $ 2.0 $ & $ 7.000 \times 10^{6} $ & $ 3.00 $ & $ 1.4 \times 10^{-2} $ & $ 9.1 \times 10^{-2} $ & $ 1.9 \times 10^{-3} $ & $ 1.7 \times 10^{-2} $ & $ 5.7 \times 10^{-1} $ \\
$ \texttt{082}\mm $ & $ 2.0 $ & $ 8.500 \times 10^{6} $ & $ 0.50 $ & $ 1.8 \times 10^{-2} $ & $ 1.1 \times 10^{-1} $ & $ 4.6 \times 10^{-3} $ & $ 1.3 \times 10^{-2} $ & $ 3.9 \times 10^{-1} $ \\
$ \texttt{083}\dd $ & $ 2.0 $ & $ 8.500 \times 10^{6} $ & $ 2.00 $ & $ 1.8 \times 10^{-2} $ & $ 1.1 \times 10^{-1} $ & $ 2.9 \times 10^{-3} $ & $ 1.9 \times 10^{-2} $ & $ 7.5 \times 10^{-1} $ \\
$ \texttt{084}\mm $ & $ 2.0 $ & $ 1.000 \times 10^{7} $ & $ 0.50 $ & $ 2.2 \times 10^{-2} $ & $ 1.2 \times 10^{-1} $ & $ 7.3 \times 10^{-3} $ & $ 1.6 \times 10^{-2} $ & $ 2.7 \times 10^{-1} $ \\
$ \texttt{085}\mm $ & $ 2.0 $ & $ 1.000 \times 10^{7} $ & $ 3.00 $ & $ 2.0 \times 10^{-2} $ & $ 1.2 \times 10^{-1} $ & $ 3.9 \times 10^{-3} $ & $ 2.1 \times 10^{-2} $ & $ 2.1 \times 10^{-1} $ \\
$ \texttt{086}\mm $ & $ 2.0 $ & $ 1.200 \times 10^{7} $ & $ 0.50 $ & $ 2.6 \times 10^{-2} $ & $ 1.4 \times 10^{-1} $ & $ 8.0 \times 10^{-3} $ & $ 1.8 \times 10^{-2} $ & $ 3.9 \times 10^{-1} $ \\
$ \texttt{087}\mm $ & $ 2.0 $ & $ 1.400 \times 10^{7} $ & $ 0.50 $ & $ 3.0 \times 10^{-2} $ & $ 1.6 \times 10^{-1} $ & $ 1.0 \times 10^{-2} $ & $ 2.1 \times 10^{-2} $ & $ 3.9 \times 10^{-1} $ \\
$ \texttt{088}\dd $ & $ 2.5 $ & $ 3.200 \times 10^{6} $ & $ 4.00 $ & $ 2.6 \times 10^{-3} $ & $ 2.0 \times 10^{-2} $ & $ 3.2 \times 10^{-4} $ & $ 3.6 \times 10^{-3} $ & $ 5.2 \times 10^{-1} $ \\
$ \texttt{089}\dd $ & $ 2.5 $ & $ 3.400 \times 10^{6} $ & $ 4.00 $ & $ 3.1 \times 10^{-3} $ & $ 2.4 \times 10^{-2} $ & $ 3.7 \times 10^{-4} $ & $ 3.8 \times 10^{-3} $ & $ 6.0 \times 10^{-1} $ \\
$ \texttt{090}\dd $ & $ 2.5 $ & $ 4.400 \times 10^{6} $ & $ 3.00 $ & $ 5.7 \times 10^{-3} $ & $ 4.5 \times 10^{-2} $ & $ 8.1 \times 10^{-4} $ & $ 7.0 \times 10^{-3} $ & $ 5.5 \times 10^{-1} $ \\
$ \texttt{091}\dd $ & $ 2.5 $ & $ 4.400 \times 10^{6} $ & $ 4.00 $ & $ 5.4 \times 10^{-3} $ & $ 4.1 \times 10^{-2} $ & $ 7.5 \times 10^{-4} $ & $ 6.8 \times 10^{-3} $ & $ 5.2 \times 10^{-1} $ \\
$ \texttt{092}\dd $ & $ 2.5 $ & $ 5.400 \times 10^{6} $ & $ 2.00 $ & $ 8.0 \times 10^{-3} $ & $ 5.9 \times 10^{-2} $ & $ 1.1 \times 10^{-3} $ & $ 9.3 \times 10^{-3} $ & $ 4.0 \times 10^{-1} $ \\
$ \texttt{093}\dd $ & $ 2.5 $ & $ 5.400 \times 10^{6} $ & $ 3.00 $ & $ 7.9 \times 10^{-3} $ & $ 5.9 \times 10^{-2} $ & $ 1.1 \times 10^{-3} $ & $ 1.0 \times 10^{-2} $ & $ 5.6 \times 10^{-1} $ \\
$ \texttt{094}\dd $ & $ 2.5 $ & $ 5.400 \times 10^{6} $ & $ 4.00 $ & $ 7.6 \times 10^{-3} $ & $ 5.6 \times 10^{-2} $ & $ 1.1 \times 10^{-3} $ & $ 1.1 \times 10^{-2} $ & $ 5.0 \times 10^{-1} $ \\
$ \texttt{095}\mm $ & $ 2.5 $ & $ 6.400 \times 10^{6} $ & $ 1.00 $ & $ 9.8 \times 10^{-3} $ & $ 7.2 \times 10^{-2} $ & $ 2.0 \times 10^{-3} $ & $ 9.0 \times 10^{-3} $ & $ 1.3 \times 10^{-1} $ \\
$ \texttt{096}\dd $ & $ 2.5 $ & $ 6.400 \times 10^{6} $ & $ 2.00 $ & $ 1.0 \times 10^{-2} $ & $ 7.1 \times 10^{-2} $ & $ 1.2 \times 10^{-3} $ & $ 1.2 \times 10^{-2} $ & $ 4.3 \times 10^{-1} $ \\
$ \texttt{097}\dd $ & $ 2.5 $ & $ 6.400 \times 10^{6} $ & $ 3.00 $ & $ 9.7 \times 10^{-3} $ & $ 6.9 \times 10^{-2} $ & $ 1.3 \times 10^{-3} $ & $ 1.4 \times 10^{-2} $ & $ 5.5 \times 10^{-1} $ \\
$ \texttt{098}\mm $ & $ 2.5 $ & $ 7.400 \times 10^{6} $ & $ 1.00 $ & $ 1.2 \times 10^{-2} $ & $ 8.3 \times 10^{-2} $ & $ 2.4 \times 10^{-3} $ & $ 1.1 \times 10^{-2} $ & $ 1.5 \times 10^{-1} $ \\
$ \texttt{099}\dd $ & $ 2.5 $ & $ 7.400 \times 10^{6} $ & $ 2.00 $ & $ 1.3 \times 10^{-2} $ & $ 8.4 \times 10^{-2} $ & $ 1.8 \times 10^{-3} $ & $ 1.5 \times 10^{-2} $ & $ 6.3 \times 10^{-1} $ \\
$ \texttt{100}\dd $ & $ 2.5 $ & $ 7.400 \times 10^{6} $ & $ 3.00 $ & $ 1.2 \times 10^{-2} $ & $ 8.4 \times 10^{-2} $ & $ 1.9 \times 10^{-3} $ & $ 1.4 \times 10^{-2} $ & $ 4.8 \times 10^{-1} $ \\
$ \texttt{101}\mm $ & $ 2.5 $ & $ 9.000 \times 10^{6} $ & $ 2.00 $ & $ 1.5 \times 10^{-2} $ & $ 9.9 \times 10^{-2} $ & $ 3.6 \times 10^{-3} $ & $ 1.5 \times 10^{-2} $ & $ 1.5 \times 10^{-1} $ \\
$ \texttt{102}\mm $ & $ 2.5 $ & $ 1.000 \times 10^{7} $ & $ 1.00 $ & $ 1.8 \times 10^{-2} $ & $ 1.3 \times 10^{-1} $ & $ 4.8 \times 10^{-3} $ & $ 1.4 \times 10^{-2} $ & $ 1.2 \times 10^{-1} $ \\
$ \texttt{103}\mm $ & $ 2.5 $ & $ 1.100 \times 10^{7} $ & $ 1.00 $ & $ 2.4 \times 10^{-2} $ & $ 1.5 \times 10^{-1} $ & $ 5.8 \times 10^{-3} $ & $ 1.6 \times 10^{-2} $ & $ 1.0 \times 10^{-1} $ \\
$ \texttt{104}\dd $ & $ 3.0 $ & $ 8.000 \times 10^{6} $ & $ 4.00 $ & $ 9.2 \times 10^{-3} $ & $ 6.9 \times 10^{-2} $ & $ 1.4 \times 10^{-3} $ & $ 1.4 \times 10^{-2} $ & $ 6.0 \times 10^{-1} $ \\
$ \texttt{105}\mm $ & $ 3.0 $ & $ 9.000 \times 10^{6} $ & $ 4.00 $ & $ 1.2 \times 10^{-2} $ & $ 8.6 \times 10^{-2} $ & $ 3.0 \times 10^{-3} $ & $ 1.4 \times 10^{-2} $ & $ 3.2 \times 10^{-1} $ \\
\end{longtable}

\label{lastpage}
\end{document}